\documentclass[useAMS,usenatbib]{mn2e}
\usepackage{graphicx,color}
\usepackage{amssymb,amsmath}
\usepackage{pstricks,float}
\usepackage[dark,timestamp]{draftcopy}

\catcode`\"=\active\let"=\"

\def\pau{P. Amaro-Seoane}
\def\rainer{R. Spurzem}
\def\emil{E. Khalisi}

\def\msol{{\rm M}_\odot}

\def\be{\begin{equation}}
\def\ee{\end{equation}}
\def\C{\begin{center}}
\def\EC{\end{center}}
\def\bfig{\begin{figure}}
\def\efig{\end{figure}}
\def\bFig{\begin{figure*}}
\def\eFig{\end{figure*}}

\def\mh{m_{\rm h}}
\def\ml{m_{\rm l}}
\def\MH{{\cal M}_{\rm h}}
\def\ML{{\cal M}_{\rm l}}
\def\MCL{{\cal M}_{\rm cl}}

\title[Mass segregation in star clusters: A comprehensive {\sc Nbody} study\\]
{A comprehensive {\sc Nbody} study of mass segregation in star clusters: Energy equipartition and escape}

\author[\emil\, \pau \, and \rainer]{\emil
  $^{1}$\thanks{E-mail: khalisi@ari.uni-heidelberg.de (EK);
    Pau.Amaro-Seoane@aei.mpg.de (PAS);
    spurzem@ari.uni-heidelberg.de (RS)}, ~\pau$^{2}$ ~and ~\rainer$^{1}$ \\
  $^{1}$Astronomisches Rechen-Institut, M"onchhofstra{\ss}e 12-14,
  Heidelberg, D-69120, Germany\\
  $^{2}$ Am Muehlenberg 1, D-14476 Potsdam, Germany}

\begin{document}

\date{Submitted to MNRAS, \today}

\pagerange{\pageref{firstpage}--\pageref{lastpage}} \pubyear{2006}

\maketitle

\label{firstpage}

\begin{abstract}
We address the dynamical evolution of an isolated self--gravitating system with
two stellar mass groups. We vary the individual ratio of the heavy to light
bodies, $\mu$ from 1.25 to 50 and alter also the fraction of the total heavy
mass $\MH$ from 5\% to 40\% of the whole cluster mass. Clean-cut properties of
the cluster dynamics are examined, like core collapse, the evolution of the
central potential, as well as escapers. We present in this work collisional
$N$-body simulations, using the high--order integrator NBODY6++ with up to
${\cal N}_{\star}=2\cdot 10^4$ particles improving the statistical significancy
of the lower--${\cal N}_{\star}$ simulations by ensemble averages.
Equipartition slows down the gravothermal contraction of the core slightly.
Beyond a critical value of $\mu \approx 2$, no equipartition can be achieved
between the different masses; the heavy component decouples and collapses.  For
the first time the critical boundary between Spitzer--stable and --unstable
systems is demonstrated in direct $N$-body models. We also present 
measurements of the Coulomb logarithm and discuss the relative importance of 
evaporation and ejection of escapers. 
\end{abstract}


\section{Introduction}

The internal evolution of a star cluster in dynamical equilibrium is governed
by its tendency to achieve a thermal velocity distribution (i.e., Maxwellian
with energy equipartition) through small changes of velocity during 2--body
encounters between stars, a phenomenon dubbed relaxation.  The relaxation time
is the average time after which a star's moving direction has been deflected by
90 degrees relative to its original orbit \citep{Spitzer87}. Relaxation
produces major changes in the structure of the cluster while keeping it in
dynamical equilibrium.

The collapse of the central core makes up an important phase and probably the
most fascinating aspect of the dynamical cluster evolution. There are three
mechanisms acting in different ways in order to achieve the collapse:
equipartition, evaporation and gravothermal instability. In a real cluster of
stars all of these processes are present, but in the idealized models, the ones
we shall describe, it is possible to isolate the specific processes and gain
some understanding of the particular effects.

\citet{Spitzer69} set about an analysis on segregation of masses in globular
clusters systems that would lead later to a broad ensemble of different
analyses and techniques. For some clusters it seemed impossible to find a
configuration in which they have dynamical and thermal equilibrium altogether.
The heavy component sink into the centre because they cede kinetic energy to
the light one on the road to equipartition. In most of the cases, equipartition
happens to be impossible, because the subsystem of massive objects becomes
self--gravitating before.  Thermal energy flows from the inner part to the
outer regions. Whereas the outer regions of the cluster do not alter
significantly their temperature, the inner regions, the core, loses heat and,
so, contracts and becomes hotter. A self gravitating system has a negative
thermic capacity.  This phenomenon has been observed in a big number of works
using different methods \citep[etc]{Henon73,Henon75,SS75a,Cohn80,MS80,Stodol82,
Takahashi93,GH94b,Takahashi95,SA96,Makino96,Quinlan96,DCLY99,JRPZ00}.  The late
phase of core collapse is the same as for a single mass model, because the
heavy components do not interact with other stars anymore.

There is an ample evidence for mass-segregation in observed clusters.
\citet{McMS94} and \citet{HH98} provided a new deep infrared observations of
the Trapezium cluster in Orion that clearly show the mass segregation in the
system, with the highest mass stars segregated into the centre of the cluster.
This is a clear-cut evidence for mass-segregation of stars more massive than 5
$\msol$ toward the cluster centre and some evidence for general mass
segregation persisting down to 1-2 $\msol$ in the Orion Nebula cluster.
\citet{RM98} study the radial structure of Praesepe and of the very young open
cluster NGC 6231. There they find evidence for mass segregation among the
cluster members and between binaries and single stars. They put it down to the
greater average mass of the multiple systems.

At this point, the question looms up whether for very young clusters mass
segregation is due to relaxation, like in our models, or rather reflects the
fact that massive stars are formed preferentially towards the centre of the
cluster, as some models predict.

To answer such questions there is a clear necessity for models that give us an
accurate description of the evolution of multi-mass models based on
direct-summation numerical schemes. 

The simplest case of a bimodal mass spectrum is a starting point to take care
of. This is a relatively good approximation if stellar black holes are the
heavy component \citep{Lee95}. Such two--mass simulations are exclusively
studied in this work.

\citet{Spitzer69} gave an analytical criterion to determine whether a
two-component system can, in principle, achieve energy equipartition or not.
According to his analysis, energy equipartition between the light and heavy
component can exist if

\be
{S}:= \Big(\frac{{\cal M}_{\rm h}}{{\cal M}_{\rm l}}\Big)
\Big(\frac{m_{\rm h}}{m_{\rm l}}\Big)^{3/2} < 0.16
\label{eq.spitzer_stab}
\ee

\noindent
Where ${\cal M}_{\rm l}$ and ${\cal M}_{\rm h}$ are the total masses
in light and heavy stars and $\ml$ and $\mh$ their individual masses,
respectively. Spitzer's work was based on many strong simplifying 
assumptions.

A number of authors has addressed the problem of thermal and dynamical
equilibrium in star clusters from a numerical point of view, with direct
$N$-body simulations \citep{PZMM00}, Monte Carlo simulations \citep{SH71b} and
with direct integration of the Fokker-Planck equation \citep{IW84,KLG98}.  As
regards the Monte Carlo scheme, recent and very detailed numerical calculations
\citep{WJR00} have suggested a different criterion,

\be
{\Lambda}:= \Big(\frac{{\cal M}_{\rm h}}{{\cal M}_{\rm l}}\Big)
\Big(\frac{m_{\rm h}}{m_{\rm l}}\Big)^{2.4} < 0.32
\label{eq.spitzer_stab_new}
\ee

The limitations inherent in this approach motivate us to embark on more
accurate models of this scenario with the help of $N$-body methods, where
Newtonian gravity is essentially treated without approximations.

In his pioneering work, \citet{Hoerner60} performed calculations with ${\cal
N}_{\star} = 16$ particles on the best computers available at that time.  Rapid
improvements in computer technology (both hard-- and software) facilitated
larger as well as more accurate calculations. The amount of $10^4$ particles
was reached by \citep{SA96}, and parallel machines and special purpose
computers do even manage fifty times more nowadays.

In this article we describe the simulation models based on these methods as
well as their initial conditions. We present the results of a wide parameter
space, which has been explored by direct $N$-body modelling for the first time.
We also extract the important parameters describing the core collapse and
equipartition of energies.

\section{Organization of the simulations: Nomenclature}

For all simulations in this work we employed a Plummer sphere model in global
virial equilibrium. The particles are treated as point masses without
softening of the gravitational force, but with regularisation of close
encounters instead. We only consider two different mass species as the most
simple approximation of a realistic mass spectrum. Since this analysis aims to
isolate the essential physical process of mass segregation, we ignore stellar
evolution, cluster rotation and a tidal field as well as primordial binaries;
binary formation occurs only during the late stage of the evolution and do not
effect our objectives.

In our notation, a {model} will be determined by its fraction of the heavy mass
component, $q := \MH/\MCL$, and the mass ratio of the individual particles,
$\mu := \mh/\ml$. The model is assigned to a capital Roman letter. Each model
consists of a number of {runs} that differ only in their random number seed
which produces different initial setup of positions and velocities of the
particles for the same distribution function. The runs are physically
equivalent. Models making up a logical unit for comparison are gathered to a
{series}, see Table \ref{tab.mass_seg1}.

\begin{table}
\begin{center}
\begin{tabular}{c|c|lc|l}
Series & Distribution & $q$ & $\mu$--models & Remarks \\ \hline
   I   &   RND        & 0.1 & A~...~H & various ${\cal N}_{\star}$   \\
  II   &   INS, OUT   & 0.1 & A~...~H & $\mh$ in / out\\
 III   &   RND        & 0.05& K~...~R & ${\cal N}_{\star}=2.5\cdot 10^3$      \\
  IV   &   RND        & 0.2 & T~...~Z & ${\cal N}_{\star}=2.5\cdot 10^3$      \\
   V   &   RND        & 0.4 & T$^{\prime}$~...~Y$^{\prime}$&${\cal N}_{\star}=2.5\cdot 10^3$\\
  VI   & RND,INS,OUT  & 0.26&  20.0   & ${\cal N}_{\star}=5\cdot 10^3$\\
\end{tabular}
\end{center}
\caption{Overview of the simulations.
   A model is described by $q$ and $\mu$ and assigned to a capital
   Roman letter.  RND = random setup; INS = all heavy masses placed
   inside; OUT = all heavy masses outside. As regards II, $\mh$ was
   inside and ouside. The series VI with ${\cal N}_{\star}=5\cdot
   10^3$ can be envisaged as a model for the Orion Nebula
\label{tab.mass_seg1}
}
\end{table}

After the work by \citet{IW84}, we concentrate on $q = 0.1$ (Series I), since
this is the value which had the fastest evolution, as they proved, and study
the evolution for a wide range of $\mu$'s (from 1 to 50). This choice is guided
by observations in youngest star clusters like the Trapezium in the Orion
Nebula Cloud, where a mass range of $\approx$ 0.1--50 M$_{\odot}$ is found
\citep{HH98}. This cluster also exhibits clear mass segregation that cannot be
explained by the simple theory of ``general'' mass segregation driven by
two--body relaxation. Given the extreme youth of the stars with highest mass, a
primordial segregation has been suggested, in which they were formed at
locations close to the dense centre \citep{BD98}.

We examine the time scales for a random setup of particles (RND) by assigning a
mass $\ml$ and $\mh$ to each body. For $\mu$ is a relative quantity, we fix
$\ml$ to unity, and vary $\mh$ in steps that are given in the top row of Table
\ref{tab.mass_seg2}. The two mass populations have ${\cal N}_{\rm l}$ and
${\cal N}_{\rm h}$ members (we will elaborate on this ahead), whose spatial
coordinates whose positions and velocities are picked up randomly according to
a Plummer sphere \citep{AHW74}.

In Series II, we place the heavy particles completely either in the centre
(INS) or in the outskirts (OUT) and compare the evolutionary patterns with the
random setup of Series I. In Series III, IV and V we alter the fraction of
heavy masses, $q$, to test for this parameter.  The full parameter space is
graphically illustrated in Figure \ref{fig:phdmodels}. Additionaly, we
performed simulations of a special configuration, which is related to a mass
ratio in the Orion Nebula Cloud (Series VI).


\bFig
  \resizebox{12cm}{!}{%
          \includegraphics{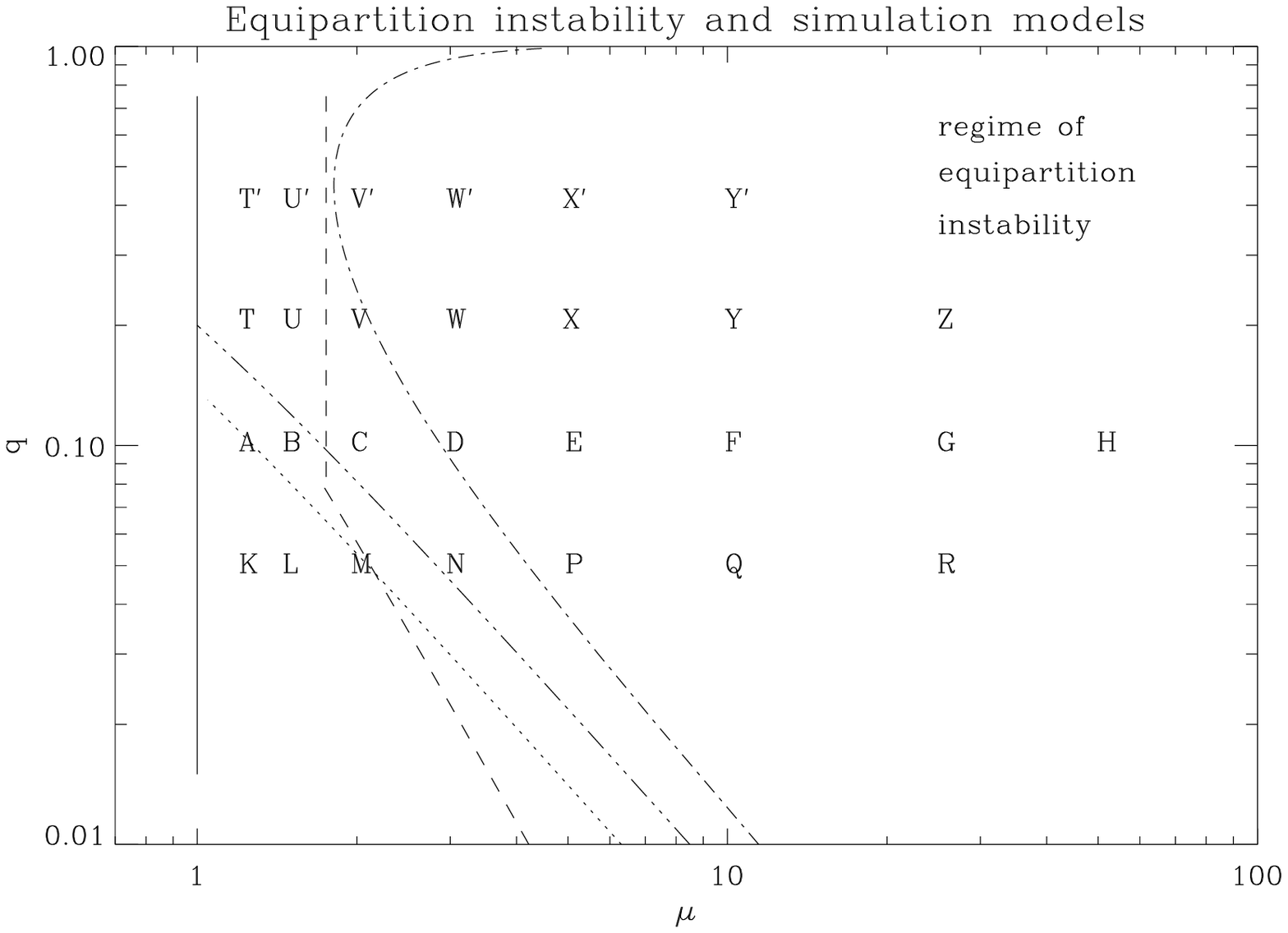}%
        }
  \hfill \parbox[b]{55mm}{%
\caption{Parameter space of the models examined in this
           work. The position of the Roman letters indicate a model
           determined by its $q$ and $\mu$.  The dotted line is the
           boundary of the equipartition stable region after
           \citet{Spitzer69}, the
           dashed--dotted line is the stability criterion after
           \citet{LF78}, the dot--dot--dot--dashed line is the
           prediction by \citet{IW84} and the dashed line is the
           empirically proposed condition by \citet{WJR00}}
\label{fig:phdmodels}
}
\eFig

In order to reduce the statistical noise, we performed a large number of runs
and averaged the data set into an ensemble--model.  The statistical quality of
such an ensemble--averaging is comparable with one single calculation
containing the full particle set of all runs \citep{GH94a}. This means that,
since the fiducial model is a cluster with ${\cal N}_{\star}=2.5 \cdot 10^3$
particles, we carried out 20 runs and the whole set containing ${\cal
N}_{\star} = 5\cdot 10^4$ has as little noise as a high-${\cal N}_{\star}$
model. In Series I, we also performed 50 runs for ${\cal N}_{\star} = 10^3$,
and 10 runs for ${\cal N}_{\star} = 5 \cdot 10^3$. Additional models containing
$10^4$ particles (4 runs) and $2\cdot 10^4$ particles (1 run) were also
performed. 

\subsection{Parameters of two--mass models} \label{ch:params}

In order to describe the physics of our models, we have to define a
set of three parameters:
\begin{eqnarray}
{\cal N}_{\star} = {\cal N}_{\rm l} + {\cal N}_{\rm h} ,\hspace{1cm}
\mu = \mh/\ml, \hspace{1cm} q = \MH/\MCL. \label{eq:mu}
\end{eqnarray}
Since the number of the force calculations per crossing time scales with ${\cal
N}_{\star}^2$, computational efforts restrict the choice of ${\cal
N}_{\star}$. We can express ${\cal N}_{\rm l}$ as follows,
\begin{eqnarray}
q & = & \frac{\MH}{\ML + \MH} = \frac{\mh {\cal N}_{\rm h}}{\ml {\cal
N}_{\rm l} + \mh {\cal N}_{\rm h}} \nonumber \\ & = & \frac{\ml \mu
{\cal N}_{\rm h}}{\ml {\cal N}_{\rm l} + \ml \mu {\cal N}_{\rm h}}
\label{eq:q} \\ & = & \frac{\mu ({\cal N}_{\star} - {\cal N}_{\rm
l})}{{\cal N}_{\rm l} + \mu ({\cal N}_{\star} - {\cal N}_{\rm
l})}. \nonumber
\end{eqnarray}
The absolute number of light particles is then
\begin{eqnarray}
   {\cal N}_{\rm l} = \frac{(1-q) \mu {\cal N}_{\star}}{q - \mu q +
   \mu}.  \label{eq:n_1}
\end{eqnarray}

Some authors (e.g. \citealt{IW84}) define a slightly different parameter
$\hat{q} := \MH/\ML$. With this definiton, equation (\ref{eq:n_1}) turns out to
be ${\cal N}_{\rm l} = \mu {\cal N}_{\star} / (\hat{q} + \mu)$.  In the
following, we employ the notation of \citet{ST95}, as defined in (\ref{eq:mu}).
The absolute numbers of the heavy particles, ${\cal N}_{\rm h} = {\cal
N}_{\star} - {\cal N}_{\rm l}$, are shown in Table \ref{tab.mass_seg2} for the
models of Series I ($q=0.1$). $\mu$ is given in brackets on top of each column.

\begin{table*}
\begin{center}
\begin{tabular}{|l||*{7}{r|}r}
\hline
 ${\cal N}_{\star}$     &    A (1.25)  &    B (1.5)    &    C (2.0)
         &    D (3.0)   &    E (5.0)    &    F (10.0)
         &    G (25.0)  &    H (50.0)   \\ \hline
$10^3$         &      82      &       69      &      53
               &      36      &       22      &      11
               &       4      &       --      \\
$2.5\cdot 10^3$&     204      &      172      &     132
               &      89      &       54      &      27
               &      11      &        6      \\
$5 \cdot 10^3$ &     408      &      345      &     263
               &     179      &      109      &      55
               &      22      &       11      \\
$10^4$         &     816      &      690      &     526
               &     357      &      217      &     110
               &      44      &       22      \\
$2\cdot 10^4$  &    1633      &     1379      &    1053
               &     714      &      435      &     220
               &      88      &       44       \\ 
\hline
\end{tabular}
\end{center}
\caption{Absolute numbers of heavy stars for models of Series I. 
Next to each model we give the value for $\mu$ in brackets
\label{tab.mass_seg2}
}
\end{table*}

A different way of fixing the mass ratio $\mu$
is by means of the average mass of the stars:
\begin{eqnarray}
   \tilde{\mu}_i = m_i / \langle m \rangle,
\end{eqnarray}
where
$\langle m \rangle = \MCL / {\cal N}_{\star}$
and $m_i$ is the $i$-th component in a multi--mass
cluster.
With $q_i = {\cal M}_i/\MCL$, we have for the general
case of $k$ different mass components
\begin{eqnarray}
   {\cal N}_i = \frac{q_i {\cal N}_{\star}}{\tilde{\mu}_i}
       = \frac{\langle m \rangle q_i {\cal N}_{\star}}{m_i}
       = \frac{{\cal M}_i}{m_i}
\end{eqnarray}
particles in the $i$--th mass bin.

The advantage of this expression lies in the simpler handling, if more
than two masses are present.  In the equation (\ref{eq:n_1}), $(k-1)!$
parameters of $m_i / m_j$ would be necessary for $k$ mass components,
while employing $\langle m \rangle$ reduces the amount of the a priori
definitions of $m_i$'s to $k-1$.

\section{Results for $\MH/\MCL=0.1$}

In this section we deal with the situation in which the fraction of heavy stars
makes up 10\% of the whole cluster mass.  We investigate the physical processes
occurring from a random initial distribution and compare them with previous
literature on this subject. The essential facts about the evolution are
qualitatively visible in Figure \ref{fig:rnd_ens}. The core radius shrinks with
time, and the cluster collapses under its self--gravity. The time scale for the
collapse is the shorter the larger the mass ratio $\mu$ between heavy and
light stars is.  The
formation of binaries in the core stops the collapse and allows its
re-expansion. We focus our attention on the variations of the evolutionary
processes for different values of $\mu$.

\bfig
\C
\resizebox{\hsize}{!}{\includegraphics
	{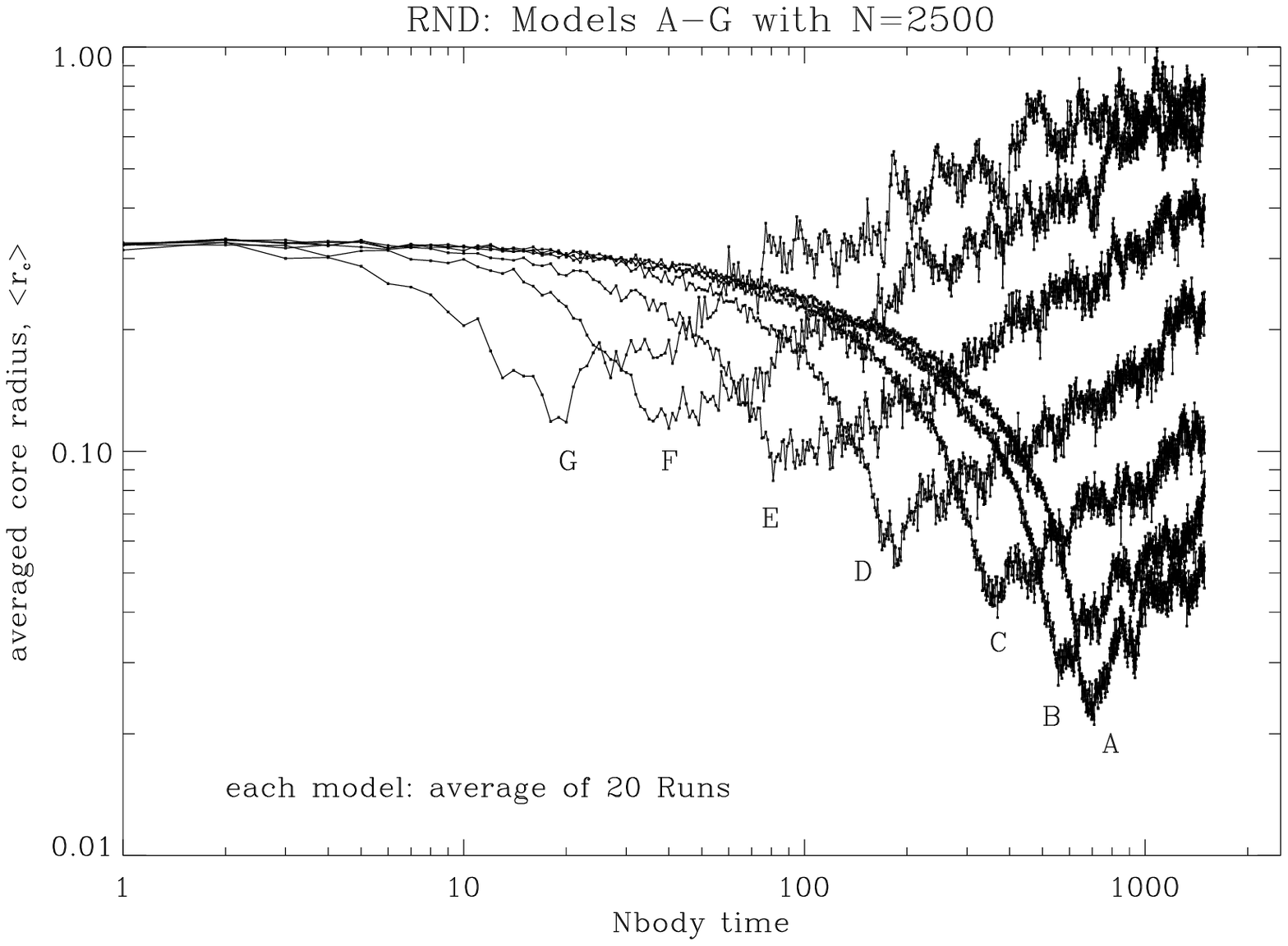}}
\caption{Comparison of the shrinking core radius in the
           models A--G of Series I given in {\sc Nbody} units
           (see Appendix \ref{ch:nbtime})
\label{fig:rnd_ens} }
\EC
\efig

\subsection{Data of ensemble averages}  \label{ch:ensembles}

We vary $\mu$ from 1.25 to 50.0 and obtain the core collapse time, $t_{\rm
cc}$, for each run by considering two values: first, the moment of the minimum
core radius, $t\,({\rm r_c|_{\rm min}})$ and, second, the deepest central
potential, $t\,({\rm \Phi|_{\rm min}})$.  The output data was written for each
$N$-body time unit. Because of large fluctuations between two subsequent data
points, we applied a ``sliding average'' over $r_{\rm c}$ and $\Phi$, based on
the following algorithm

\begin{eqnarray} {\cal R}_i =\frac{1}{w}\sum_{j=0}^{w-1}{\cal A}_{i+j-w/2} ,
\end{eqnarray}
where ${\cal R}_i$ is the resulting value of the original variable ${\cal A}_i$
averaged among $w$ neighbouring data points. The choice of $w = 5$ turned out
to be best fitting one.

\bfig
\C
\resizebox{\hsize}{!}{\includegraphics
	{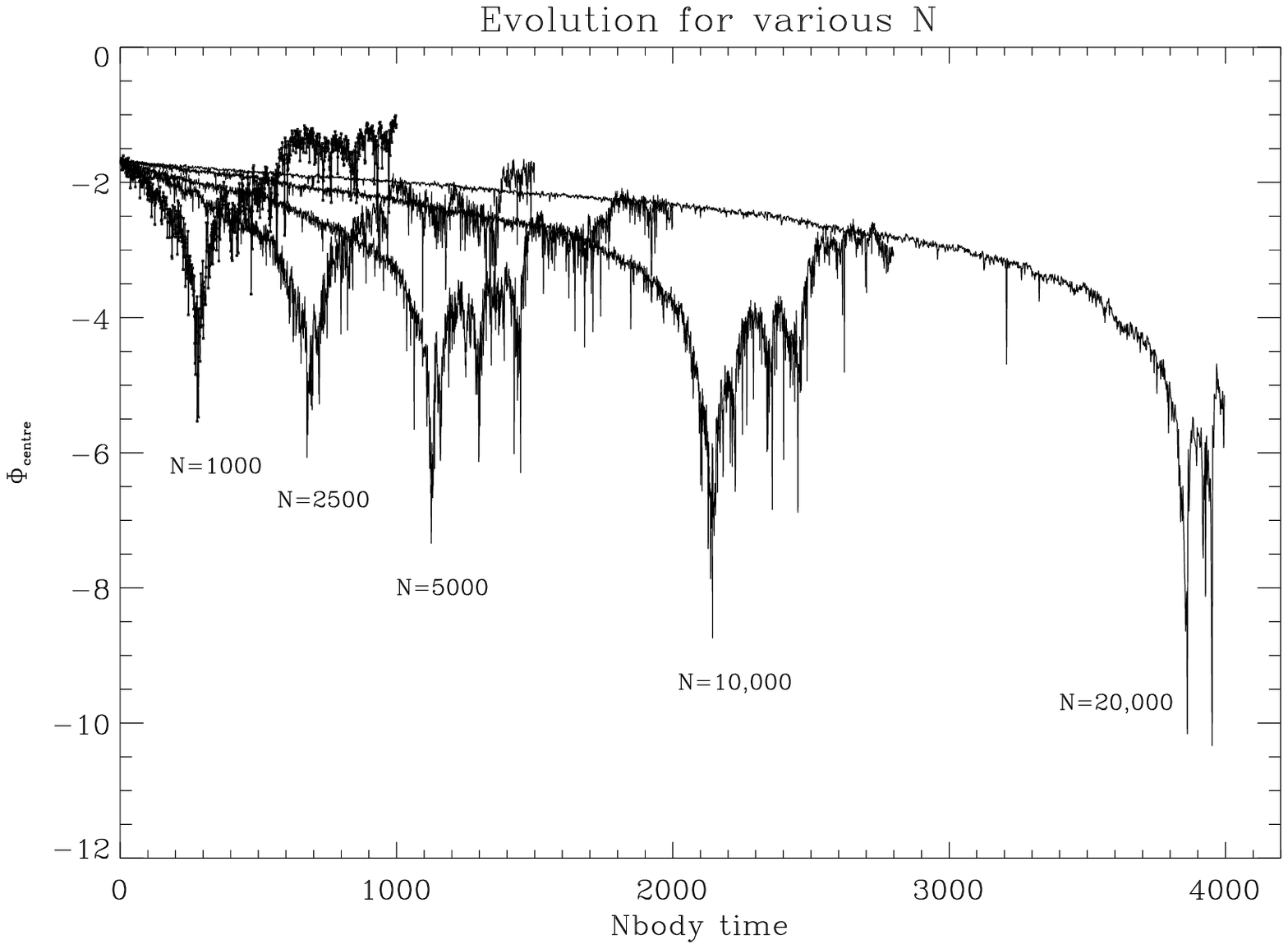}}
\caption{Evolution of the central potential of selected
           runs of Model A for ${\cal N}_{\star} = 10^3, 2.5\cdot
           10^3, 5\cdot 10^3, 10^4~{\rm and}~ 2\cdot 10^4$ particles
\label{fig:phiforn} }
\EC
\efig

The average of the times found from the minima of $r_{\rm c}$ and $\Phi$ is
defined to be the core collapse time of the run:
\begin{eqnarray}
t_{\rm cc} = \frac{ t_{r_{\rm c}} + t_{\Phi} }{2}. \label{eq:tccexp}
\end{eqnarray}

\bfig
\C
\resizebox{\hsize}{!}{\includegraphics[bb=78 370 291 602,clip]
	{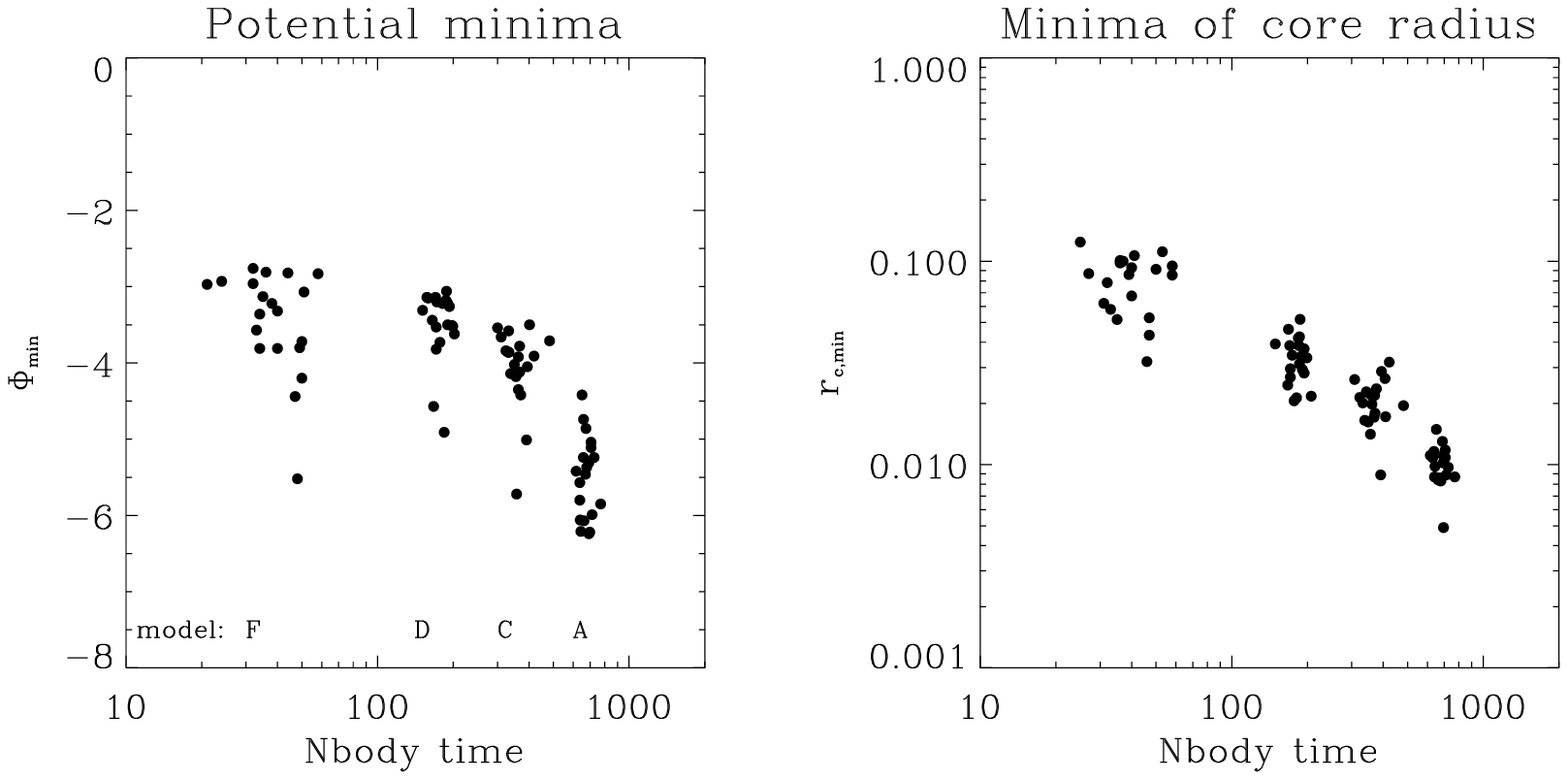}}
\resizebox{\hsize}{!}{\includegraphics[bb=319 370 547 602,clip]
	{grafiques/fmodelcomp2.ps}}
\caption{Models A, C, D, and F of Series I with ${\cal N}_{\star}=2500$.
           Each point represents the minimum value of the
           central potential $\Phi_{\rm c}$ (upper panel) and
           core radius $r_{\rm c}$ (lower panel) at the time
           when these parameters attain their minimum.
           The average values for all models are listed
           in the Appendix
\label{fig:modelcomp2} }
\EC
\efig

Figure \ref{fig:modelcomp2} illustrates the distribution of the data
points of $r_{\rm c}$ and $\Phi_{\rm min}$ at their corresponding core
collapse times for four selected models. The logarithmical time--axis
was chosen to show the relative scatter between the point clouds.
When looking at a particular run, the core collapse times are likely
to show a significant discrepancy between the determination from the
potential or from the core, i.e.\ $t_{\rm \Phi}$ does not necessarily
correspond to $t_{\rm r_c}$, but the averages of both, $\langle t_{\rm
r_c}\rangle$ and $\langle t_{\rm \Phi}\rangle$, yields a good
concordance for the whole model. Thus, $\langle t_{\rm cc}\rangle$ is a
good value to characterize the core collapse time of the model,

\begin{eqnarray}
\langle t_{\rm cc}\rangle
    = \frac{\sum \left[ (t_{\rm r_{\rm c}} + t_{\Phi})/2 \right] }
    {{\cal N}_{\rm runs}} = \frac{\langle t_{r_{\rm c}}\rangle +
    \langle t_{\Phi}\rangle }{2}.
\end{eqnarray}
In this last equation ${\cal N}_{\rm runs}$ is the number of runs
and we obtain an equality between the left-hand and right-hand side
by linearity.

As we can see in Fig.\,(\ref{fig:modelcomp2}), the relative variance of the
core collapse times increases for smaller values of $\langle t_{\rm
cc}\rangle$. 
%
The mean values for the eight models of our Series I are summarized in Appendix
A. The errors given there are the standard deviation from the runs' mean
$t_{\rm cc}$, divided by the square root of the number of the runs $\sqrt{{\cal
N}_{\rm runs}}$.  The error of $\langle t_{\rm cc}\rangle$ is roughly 2--5\%
for most of the models, consistent with the relative errors determined by
Spurzem \& Aarseth (1996) and the half--mass evaporation times by Baumgardt
(2001).

\section{Evolution of the core radius}

The evolution of the core radii for the models A--G is plotted in Figure
\ref{fig:rnd_ens}. Each curve is an ensemble average of 20 runs.  For models
with $\mu$ approaching unity, the core radius shrinks as in an equal--mass
system (at late collapse times, to the right). A linear time scale (not shown
here) suggests that the collapse phase sets in when the core radius has
contracted to about 25\% of its initial radius. This is in agreement with the
results of \citet{GH94a}. For high $\mu$'s, this happens from
the start of the simulations. The rapid contraction of $r_{\rm c}$ is due to
the very massive stars falling quickly to the centre. The
contraction stops at higher $r_{\rm c}$--values than for the low--$\mu$ models,
and a quick expansion of the core follows.

The behaviour at the moment of core bounce is illustrated in Figure
\ref{fig:tccmu2}, where the minima of the potential and the core radius are
plotted versus $\mu$. For a fixed ${\cal N}_{\star}$ and values of $\mu$
between 3-10, the core collapse is carried out by approximately the same number
of particles, but they do not draw together as close as for the equal--mass
case or very small $\mu$'s.  Therefore, the density and the central potential
are less deep than for $\mu \lesssim 2$, and the minimum of the core radius is
not so profound. At high $\mu$'s, the effect is reversed: Even some few heavy
components are massive enough to deepen the potential. The collapse itself is
less distinct, as seen from the shallower $r_{\rm c}$ in the right panel. In
this range of values for $\mu$, the depth of the potential results from the
combination of these two effects: Heavy masses build up a strong gravitational
field, but their kinetic motion does not allow a long--lasting vicinity.
Equipartition becomes then impossible. The differences of the depths for
{various} ${\cal N}_{\star}$ are discussed in section \ref{ch:largen}.

\bfig
\C
\resizebox{\hsize}{!}{\includegraphics[bb=70 370 300 599,clip]
	{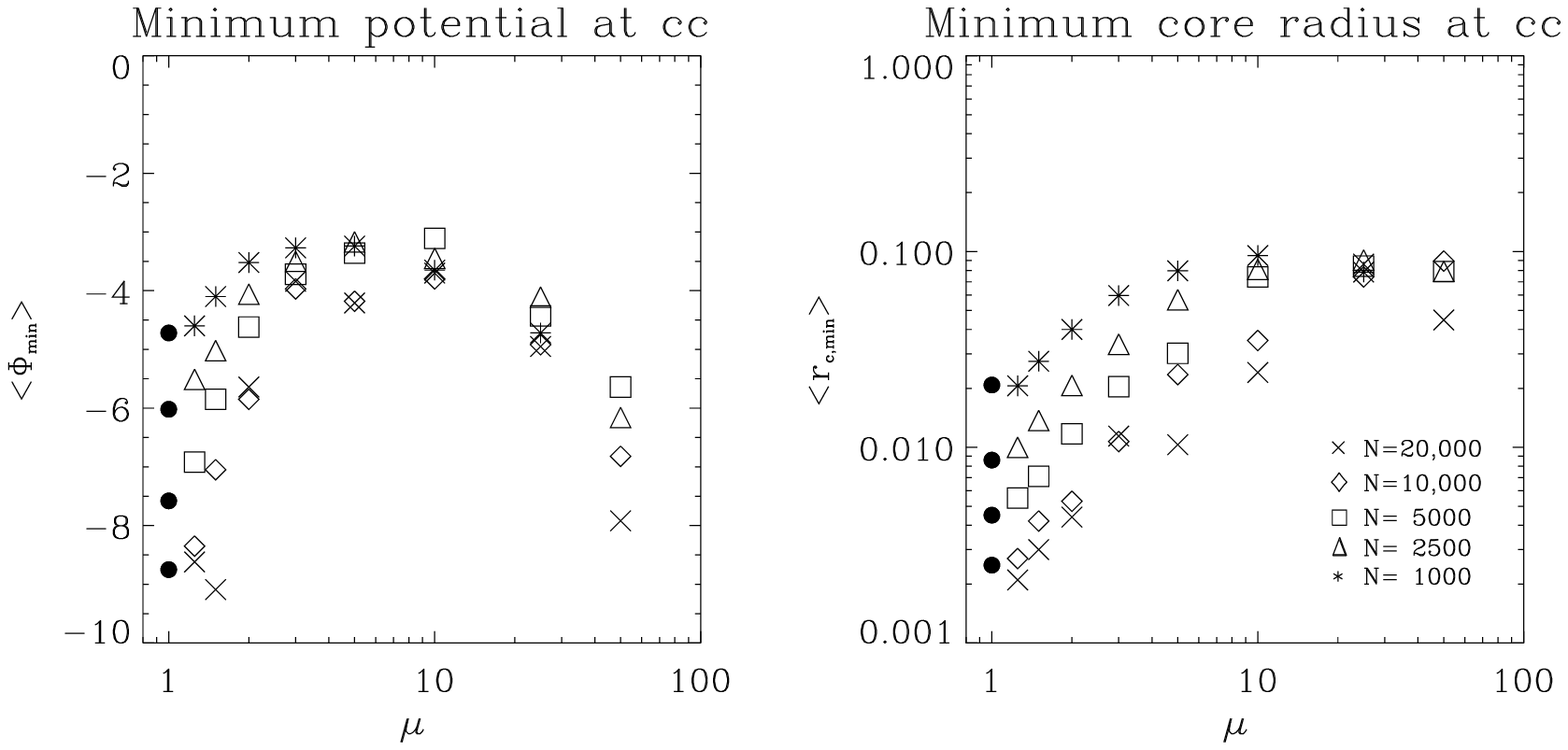}} \vspace{0.2cm}
\resizebox{\hsize}{!}{\includegraphics[bb=317 370 557 599,clip]
	{grafiques/ftccmu2.ps}}
\caption{Minima of potential (upper panel) and core radius
           (lower panel) at the moment of core bounce.
           The collapse is carried out by a less number of
           core particles when $\mu$ rises. The filled circles
           correspond to the single-mass case.
           See text for explanations in more detail
\label{fig:tccmu2} }
\EC
\efig

For a system in which the central object has a small mass and the energy
production is confined to a small central volume we have that the core-radius
$r_c$ should follow an expansion proportional to the power-law of $t$ and
$t_{\rm cc}$ \citep{Henon65,Shapiro77,McMLC81,Goodman84},

\begin{eqnarray}
  r_{\rm c} \propto (t-t_{\rm cc})^{2/3}. \label{eq:postcc}
\end{eqnarray}

%
%
We measured the slopes of the expanding $r_{\rm c}$'s in Figure
\ref{fig:rnd_ens} by fitting two straight lines embracing the fluctuating data
of each model to construct an upper and a lower margin. These lines give two
independent measurements for the slope and are presented in Table
\ref{tab.mass_seg3}, and they show a decreasing trend for higher $\mu$.

\begin{table}
\begin{tabular}{l|*{7}c}
slope   &   A   &   B   &   C   &   D   &   E   &   F   &   G   \\ \hline
up      & 0.631 & 0.666 & 0.680 & 0.525 & 0.522 & 0.490 & 0.420 \\
low     & 0.940 & 0.864 & 0.662 & 0.615 & 0.524 & 0.517 & 0.445 \\ \hline
mean    & 0.7855& 0.765 & 0.671 & 0.570 & 0.523 & 0.5035& 0.4325
\end{tabular}
\caption{Measured exponent for core radius expansion in 
  the post--collapse phase. See text for further explanation.
\label{tab.mass_seg3}
}
\end{table}

The lower lines seem to be steeper than the upper ones for the most of the
models. On the other hand, the simulations reach different stages of the
post--collapse evolution, and the expanding branches exhibit different lengths.
Especially, the low--$\mu$ models (A and B) are not far advanced for precise
measurements, while the high--$\mu$ models (F and G) appear distorted about the
time of collapse such that the onset of the self--similar expansion is
difficult to find (see also \citealt{GH94b}). From the theoretical point of
view, there is no argument for a different behaviour of the core expansion when
unequal masses are present.

\section{Core collapse times}  \label{ch:cctimes}

As we explained in the introduction, two--body relaxation causes star clusters
to redistribute the thermal energy among stars. Since this kind of heat
transfer acts on the relaxation time scale, a core collapse is similarly ensued
in gravitationally unstable systems. 

The core collapse time is best studied numerically. For equal--mass models it
ranges about $t_{\rm cc} \approx 330\, t_{\rm rc}(0)$ during the self-similar
phase. or about 12--19 half--mass relaxation times \citep{Quinlan96}. In this
article, Quinlan gives a time scale of $15.7\, t_{\rm rh}$ for an isolated
cluster, if an isotropic velocity distribution is assumed.  \citet{Takahashi95}
also modelled Plummer spheres but for an anisotropic case, and determined the
collapse time to about 17.6$\, t_{\rm rh}$. Other authors find similar factors,
and we shall adopt
 
\begin{eqnarray} t_{\rm cc} \approx 17.5\, t_{\rm rh}  .   \label{eq:tcctheo}
\end{eqnarray}

Though the values are used in most studies of the core collapse, they are a
poor guide for clusters with a mass spectrum, e.g.\ globular clusters have
central relaxation times that are typically ten, sometimes a hundred times
shorter than their half--mass relaxation times \citep{Quinlan96,GurkanEtAl04}.

So far, the core collapse time is only found empirically from a large number of
numerical simulations, for there exists no analytical theory which would
predict it a priori from cluster properties, e.g.\ the star number, IMF, or
concentration parameters.

It has been suggested that the nucleus consisting of two mass--components
collapses in a time shorter than the equal--mass cluster by a factor of $1/\mu$
\citep{FregeauEtAl02}. In broad terms this is so because the equipartition time
is shortened according to the time-scale for equipartition \citep{Spitzer69}.
The upper panel of Figure \ref{fig:tccmu} shows detailed calculations in a wide
$\mu$--range of this. The mean times, $\langle t_{\rm cc}\rangle$, have been
plotted versus $\mu$ for the complete sample of our models in Series I. 
\bfig
\C
\resizebox{0.9\hsize}{!}{\includegraphics[bb=61 370 301 602,clip]
	{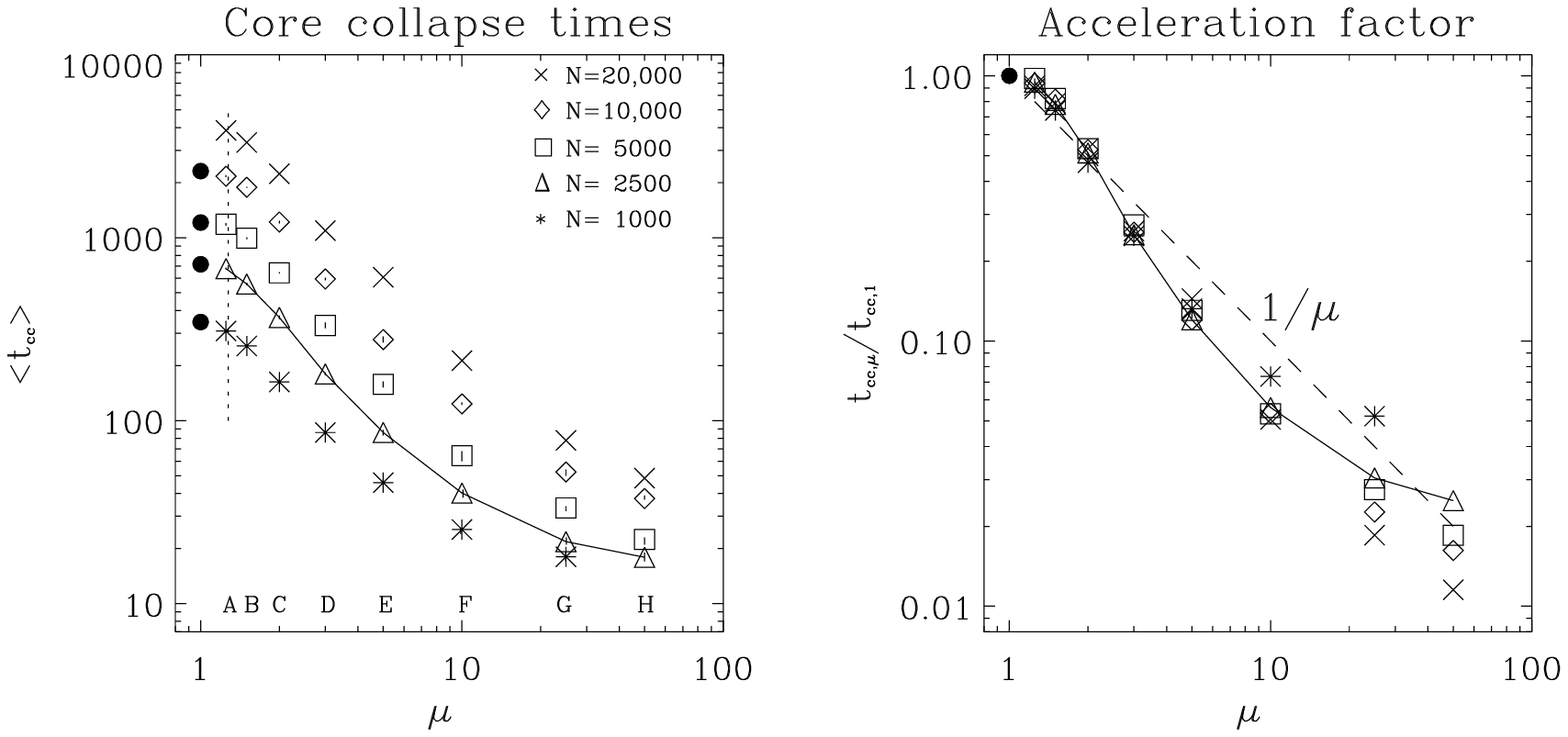}}
\resizebox{0.9\hsize}{!}{\includegraphics[bb=324 370 553 602,clip]
	{grafiques/ftccmu.ps}}
\caption{Upper panel: Core collapse times of the models
           in Series I.  The times shorten non--linearly with
           increasing $\mu$.  The dotted line at $\mu=1.275$ is the
           equipartition criterion given by
           \citet[Eq. 3-54]{Spitzer69} for a value of $\chi_{\rm max}
           = 0.16$ and $q=0.1$.  The core collapse times of the
           corresponding equal--mass clusters are indicated by filled
           dots.  The error bars are smaller than the symbols. In the
           lower panel all collapse times are normalised to $t_{\rm
           cc,1}$ (the core collapse time for the single--mass case)
           The simulations for ${\cal N}_{\star}=2.5\cdot 10^3$ are 
           connected with a solid line for clarity. The circles at left 
           are the collapse times for an equal--mass cluster according 
           to eq.~(\ref{eq:tcctheo}), and the dotted line is the stability 
           boundary by Spitzer with $\chi_{\rm max} = 0.16$
(\citealt{Spitzer87}, Eq. 3-55).
\label{fig:tccmu} }
\EC
\efig

In the upper panel, the core collapse times converge smoothly to the collapse
time for an equal--mass cluster, $t_{\rm cc,1}$, as $\mu$ approaches unity.
Between $\mu \approx 2$ and 10, a somewhat linear decline is visible, but far
beyond the stability boundary, run towards a constant value, since binaries are
likely to have more importance the larger $\mu$ is.  In the same Figure, in the
lower panel, we display the factor of ``collapse acceleration'' for various
$\mu$'s. This gives us a measure to which percentage a cluster with two masses
evolves faster than the single--mass case. The dashed line is the
$\mu^{-1}$--decline, i.e.\ a cluster with two mass--species having a ratio
$\mu$ would collapse $\mu$-times earlier than its single--mass equivalent
according to

\begin{eqnarray}
  t_{\rm cc,\mu} \propto \frac{1}{\mu}t_{\rm cc,1}.
\end{eqnarray}

At $\mu=2$, the decoupling of the equipartition--based instability and the
gravothermal instability seems to take place. For $\mu \longrightarrow 1$, the
evolution occurrs more slowly than the $1/\mu$--decrease, because the tendency
to equipartition drives the initial evolution and slows down a ``purely''
gravothermal collapse. This seems to suggest that as the mass difference
between $\mh$ and $\ml$ becomes less important, the system collapses like a
single--mass cluster. Beyond the critical value $\mu=2$, an early decoupling of
the two mass populations occurs.  The heavy components try to reduce their
large velocity dispersion, but they rapidly accumulate in the centre and
interact preferably with themselves.  As a consequence, equipartition is harder
and harder to achieve and the evolution proceeds only due to the redistribution
of heat within the two, almost separated components \citep{BI85}.  The light
stars evaporate out of the core and take away the thermal energy to the
outskirts, while the heavy components increase their binding energy. The latter
ones collapse like a single--mass subcluster. The release of energy transported
away and the heat transfer works effectively and leads to an accelerated
collapse.

If $\mu \gg 1$, the situation turns into a case of dynamical friction: A
significant fraction of particles are drowned into a homogeneous sea of light
stars and, like in an ordinary frictional drag, their motion suffers a
deceleration. It is instructive to see that only for large ${\cal N}_{\star}$,
the {absolute} number of heavy stars seems sufficient to maintain the linear
slope of accelerated cluster evolution. The slope follows the dashed line a bit
longer before bending towards that constant value. 

\section{Mass segregation}

The process of mass segregation for the six models A--F is illustrated in the
Figure \ref{fig:fallshell}, where the mean mass of the stars that are inside a
specified Lagrangian shell, is shown (i.e., a Lagrangian shell is the volume
between two Lagrangian radii, which contain a fixed mass fraction of the bound
stars in the system; see \citealt{GH94a}).  With the light and heavy masses
randomly distributed, each shell exhibits the same average mass at the
beginning. In the course of the evolution, the inner shells assemble the heavy
bodies, and raise the mean mass. The half--mass radius and the outer shells
lose their heavy stars rather quickly and remain below the value for the
average mass, because the light ones outnumber the heavy components
significantly. The cluster is stratified by mass.

\bFig
\C
\resizebox{\hsize}{!}{\includegraphics
	{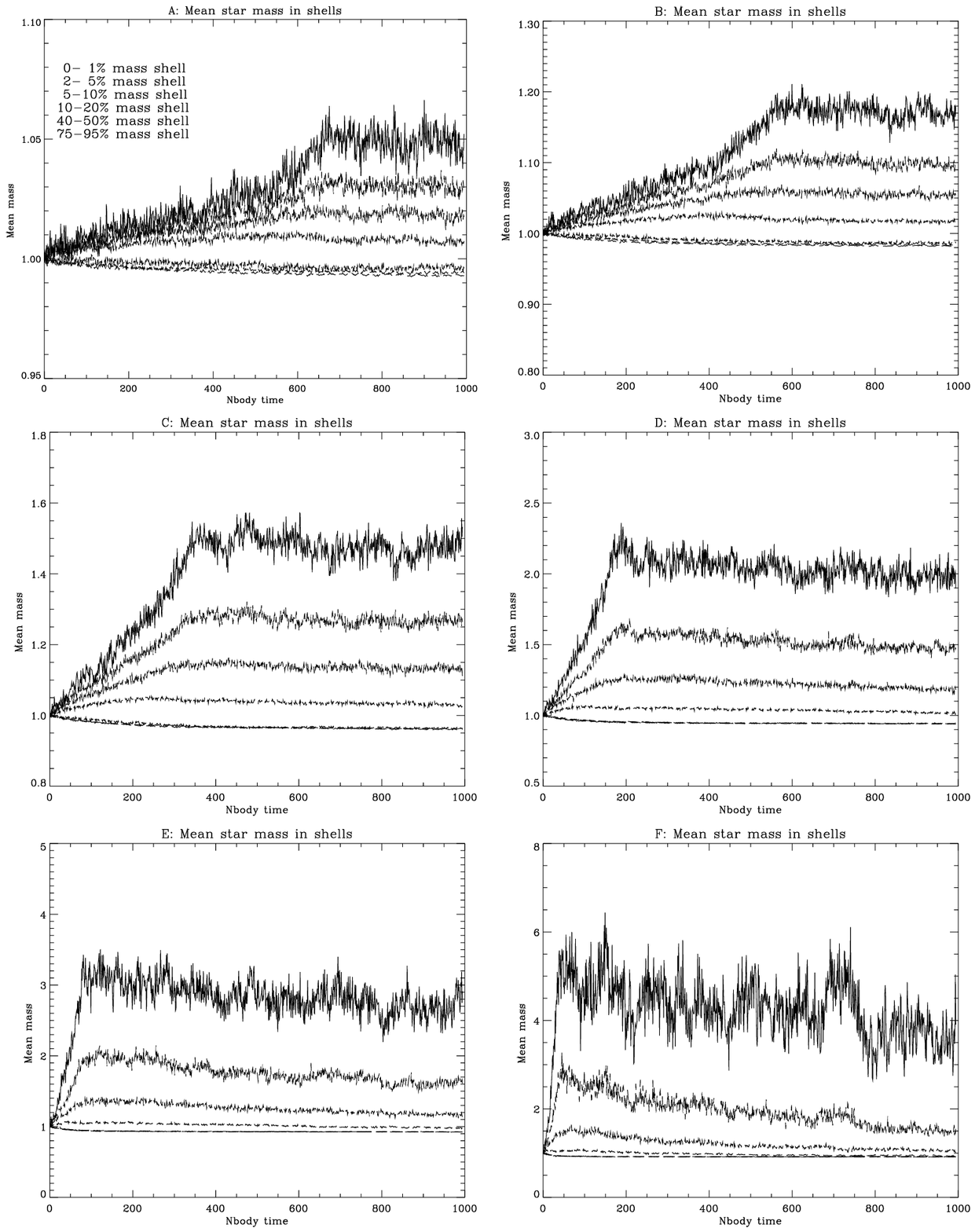}}
\caption{Average mass in Lagrangian shells for models A--F
           show the stratification of masses in the cluster.
           The segregation of heavy masses proceeds in
           agreement with the global evolution of
           the cluster.
           The mean mass is indicated
\label{fig:fallshell} }
\EC
\eFig

\bFig
  \resizebox{12cm}{!}{%
          \includegraphics{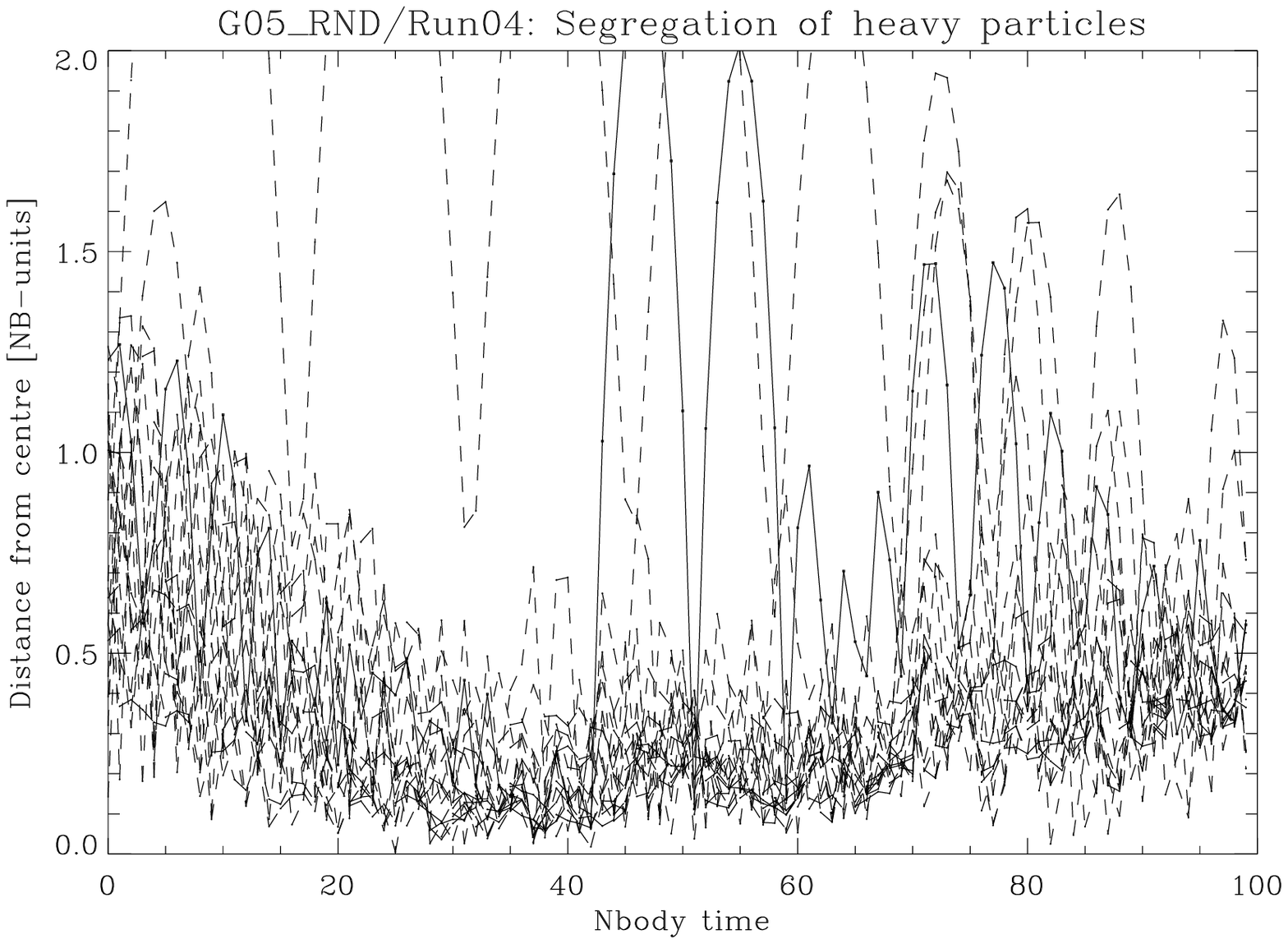}%
        }
  \hfill \parbox[b]{55mm}{%
\caption{Distance moduli of 18 heavy mass particles
           in Model G with ${\cal N}_{\star}=5\cdot 10^3$.
           Each of the particles has a mass 25 times
           larger than a light particle (orbits
           not shown).
           The core collapse of this run occurred at
           $t_{\rm cc} = 33.5$}
\label{fig:segreg2}
}
\eFig

The segregation of masses propagates simultaneously with the contraction of the
core. \citet{GH96} have already noticed the self--similarity of this process:
In the models A and B we can clearly see how the inner layers decouple stepwise
one after the other until the final stage is reached. After the collapse, the
profile does not change much; the effect has only influence on the early
evolutionary phase, not in the post--collapse. The moderate decline in the
models E and F can be explained by the heavy particles escaping: Because of
their relatively small absolute number (see Table \ref{tab.mass_seg2}), they
get ejected from the core and leave the cluster so that the surplus of small
bodies depletes the mean value in the shells. A fraction of high--mass stars in
the centre interact strongly in a few--body process and kick out each other,
and the core is gradually ``evacuated'' from the heavy stars. In our data we
find an enhanced fraction of high--mass escapers occuring immediately after the
collapse confirming this scenario; many of them exhibit enormous kinetic
energies and some even escape as bound binaries (see Sec.~\ref{ch:esc}).
However, some heavy latecomers enter the inner regions, but a balance is found
between the incoming and outgoing mass flux.  In general, the dynamical
processes in the core do not influence the properties of the cluster as a
whole.

The mean mass in the 1\%--Lagrangian shell never attains its full ``capacity'',
i.e.\ the value of $\mh$ that could principally be gained if this shell was
completely populated by heavy stars; which is only possible if the heavy stars
are 51\%. It means that there is always a number of low--mass stars entering
and leaving the innermost region that keep $\langle m\rangle_{1\%}$ at a
constant fraction of the maximally attainable level. This level is at about
0.8$\, \mh$ if $\mu$ is small, and drops to 0.5$\, \mh$ for the highest
$\mu$'s.  

Another illustration of the segregation process is given in Figure
\ref{fig:segreg2}. It shows the shrinking distance moduli of 18 heavy--mass
stars of one typical run of a model with a high mass--ratio (Model G, $\mu=25$,
Run no.\ 4, ${\cal N}_{\star}=5\cdot 10^3$); this model consists of 22 heavy
mass stars initially, but 4 stars escaped at some time before the moment for
which we show the plot and they were excluded from it. 17 dashed lines were
overlapped demonstrating how the orbits of most particles draw rapidly closer.
The core collapse occurs at $t_{\rm cc} = 33.5$.  A solid line lifts off one
particle of example that was knocked out of the chaotic region due to a close
encounter, but its kinetic energy was not sufficient enough to leave the system
and it segregated inwards again. One other particle, whose pericenter was at
0.9 $N$-body--radii, remained in the halo and did not take part in the
collapse, but it sank to intermediate distances at about $t_{\rm cc} \approx
70$.

\section{Escapers} \label{ch:esc}

The estimation of escape rates is often based on idealized models, whose
simplifying treatments sometimes lead to different results. The complexity of
this topic is reviewed by \citet{MH97}. For the results that we show in this
work, we should mention that:

\begin{itemize}
\item Theories based on diffusive or small-angle relaxation 
      phenomena yield a different escape rate than theories 
      involving individual two--body encounters;
      the former is often denoted as ``evaporation'',
      the latter is related to ``ejections'', which dominate
      for isolated clusters, like here
\item The rate of escape is not a constant, while the
      evolution of the system proceeds, even in
      the pre--collapse phase
\item An increasing concentration in the core
      as well as the growth of anisotropy tends
      to enhance the escape rate
\item Furthermore, the escape rate is strongly
      mass dependent; different mass spectra
      and segregation alter it
\item A tidal field lowers the energy threshold for escape
\item A sufficient abundance of binaries (both,
      primordial and formed) has a substantial
      effect on high--velocity escapers which take
      away energy from the system 
\end{itemize}
In view of these complications, we have to take care in the
interpretation of the data. We show in this section $N$-body
simulations investigating the rates for various particle numbers and
the variation on $\mu$, in particular.

An escape is defined by a particle having both positive energy and its distance
from the density centre exceeding a limiting radius.  We have chosen the
distance to be 20 times the half--mass radius, $r_{\rm h}$.

The particles were removed from the calculation when both conditions were
fulfilled; we shall call them ``removed escapers''. We focus here on escapers
occuring before the time of core collapse, $t_{\rm cc}$. One run contains a
number of particles which have gained positive energy but not reached the
distance for a removal yet. Such kind of particles dominate when $t_{\rm cc}$
is very short. In particular, the high--$\mu$ models collapse within a few tens
of $N$-body--time units, and a large number of particles, that are going to
escape, would be missed.  This situation resembles the ``energy cut--off''
\citep{Baumgardt01}, and we shall call these particles ``potential escapers''.
Whether some of them will be scattered back to become bound members again or
really escape, is a complex process that is out of the scope of our subject.
\citet{Baumgardt01} estimated that a fraction of 2\% of the potential escapers
might return to the system. As a first approach to the general properties of
escapers, we will assign to the number ${\cal N}_{\rm esc}$ all removed
escapers plus potential escapers at the epoch of $t_{\rm cc}$. Because of the
scatter of the core collapse times (section \ref{ch:ensembles}), each run was
checked separately for its escapers that occured before that run's individual
core collapse time --- not the ensemble's $\langle t_{\rm cc}\rangle$.
$\langle {\cal N}_{\rm esc}\rangle$ is the average among the individual ${\cal
N}_{\rm esc}$'s.

Figure \ref{fig:esc_c05} gives a typical example for the energy distribution of
all removed escapers in one run of Model C with ${\cal N}_{\star}=5\cdot 10^3$;
the potential escapers are disregarded here.  The energy is
plotted against the time of removal; it is measured in units
of $kT = {\bar m}\sigma_{0,av}^2$, where ${\bar m}$ is the average mass and
$\sigma_{0,av}$ the average 1D velocity dispersion in the core, measured 
for the initial model.
Light bodies are
represented by filled dots, heavy ones of $\mh = 2.0$ by diamonds.  The
collapse occured at $t_{\rm cc} = 643.5$ for this run.  In the beginning, the
light particles diffuse slowly from the system with small energies
(evaporational effect).  After the core collapse, the mass dependence is more
complicated, for a second class of escapers joins: high--energy particles,
whose energies are higher by two orders of magnitude.  A fair fraction of the
escapers are heavy stars.  As in the statistics for equal--masses by
\citet{GH94a}, it is natural to associate them with ejected stars that go back
to three--body interactions in the very centre.  Mass segregation has widely
finished at the time of core collapse, and interactions in the core start
depleting the high--mass population. Other runs of the same model exhibit a
similar picture of the physical scenario.


\bFig
  \resizebox{12cm}{!}{%
          \includegraphics{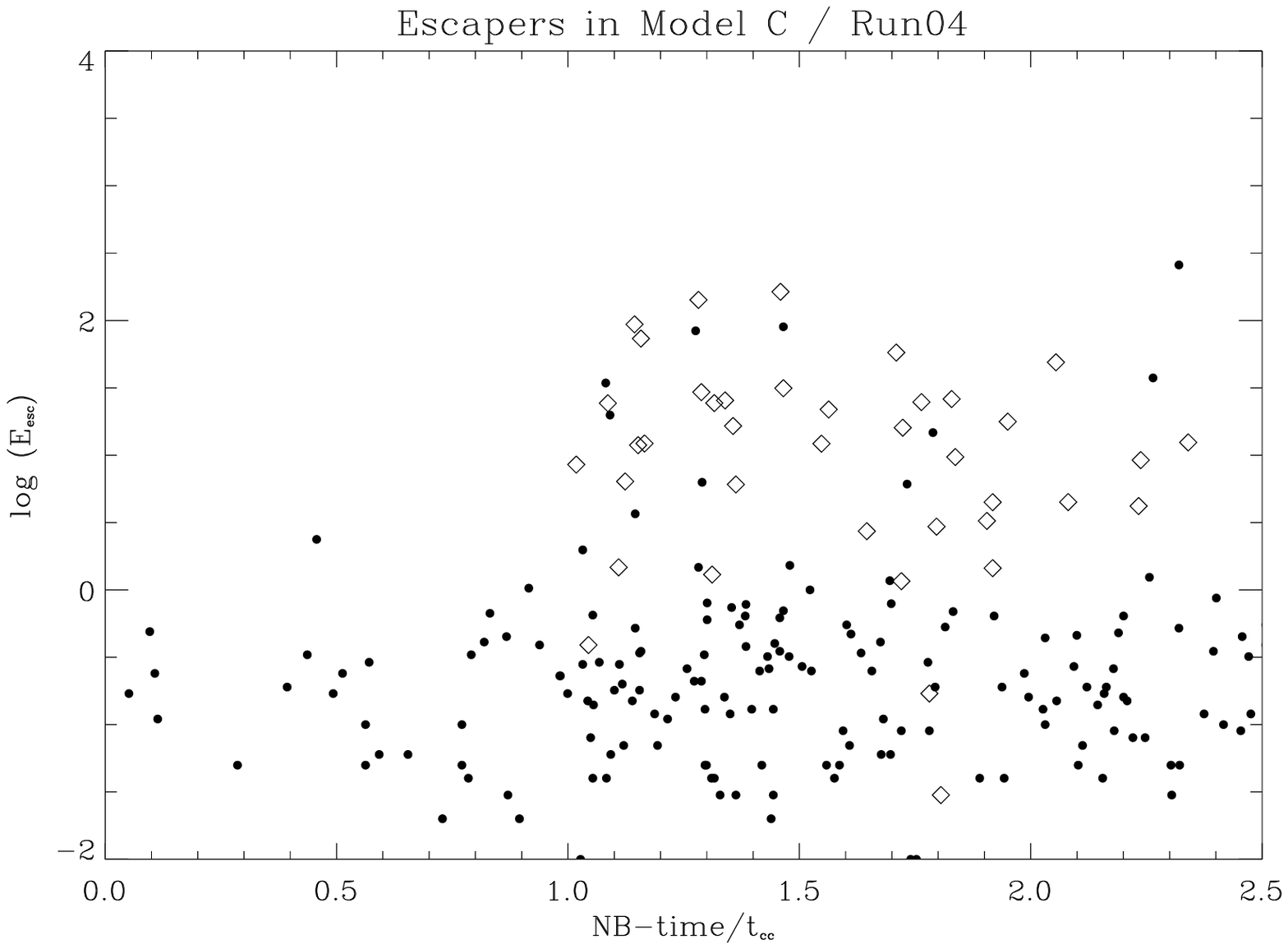}%
        }
  \hfill \parbox[b]{55mm}{%
\caption{Model C ($\mu = 2.0$), Run no.~04, with 
${\cal N}_{\star}=5\cdot 10^3$.  Energy of an escaping particle is
given in units of $kT$ and plotted against its escape time
scaled to the core collapse.  Diamonds denote heavy masses}
\label{fig:esc_c05}
}
\eFig

The analysis of the models in regard to $\mu$ reveals interesting views on the
escape mechanism. A summary is shown in Figure \ref{fig:escrate}, and the
legend for the symbols is given in panel (e). Panel (a) gives the fraction of
escaped stars, $\langle {\cal N}_{\rm esc}\rangle/{\cal N}_{\star}$. In a
single--mass cluster ($\mu=1$), about 2.5\% of the stars leave the system
before it collapses. When introducing a second mass, this fraction drops to
0.2--0.5\% until $\mu \approx 3$. The reason is that the rate of escape (panel
e) is nearly constant for small $\mu$, but the shorter collapse times cause a
smaller progress of the escape mechanism, and thus a smaller ${\cal N}_{\rm
esc}$.  When massive bodies $\mh \gg \ml$ are present, a larger fraction of
stars receives positive energy and turns into potential escapers. 

\bFig
\C
\resizebox{0.9\hsize}{!}{\includegraphics
	{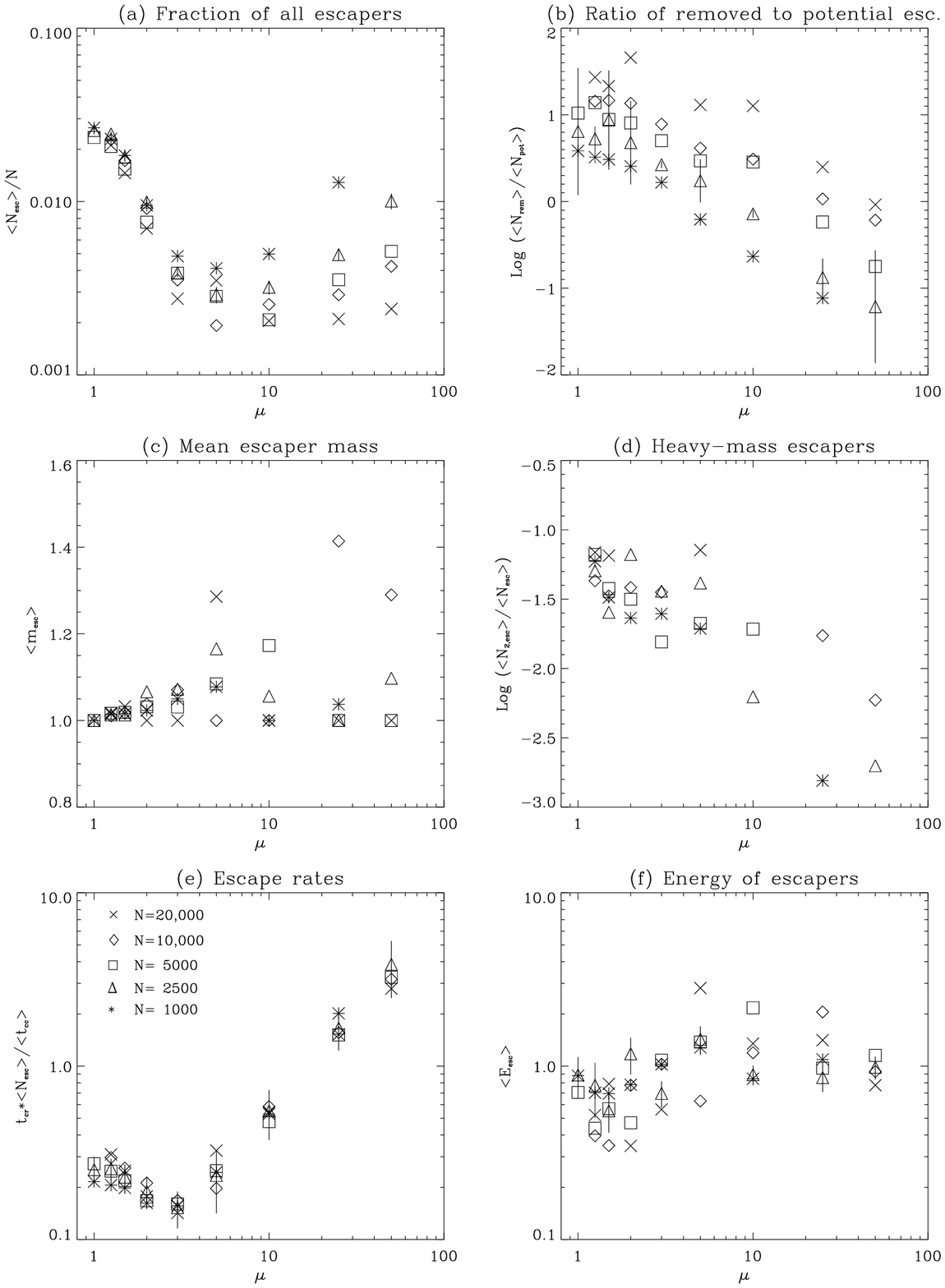}}
\caption{Summary of escaper data for the different models.
           (a) Fractions of removed escapers before
               the core collapse plus potential escapers
               at $t_{\rm cc}$.
           (b) Ratio of removed escapers to  potential
               escapers.
           (c) Mean mass of escapers.
           (d) Relative number of $\mh$--escapers among
               all escaping particles.
           (e) Escape rates.
           (f) Mean energy of escapers in $kT$.
The error bars are usually omitted, though they are given for the ${\cal
N}_{\star}=2.5 \cdot 10^3$ set (triangles) except in the panels (c) and (d);
the error bars are almost invisible in (c), and larger than the panel size in
(d), because the number of high--mass escapers varies a lot among one model.
\label{fig:escrate} }
\EC
\eFig

Panel (b) gives the ratio of the removed escapers to the potential escapers as
described above, $\langle N_{\rm rem}\rangle /\langle {\cal N}_{\rm
pot}\rangle$. For $\mu$ close to the equal--mass case, the core collapse time
is large.  Thus, the accumulated number of removed particles is larger by
$\approx$ 10 times than the number of potential escapers at the moment of
$t_{\rm cc}$. For high $\mu$'s, removed particles are scarce, while potential
escapers did not have time to cross the cluster and turn into removed ones.
Therefore, the potential escapers make the overwhelming majority. The slope of
the decrease is $-1$ , and it is similar to the shortening of $t_{\rm cc}$
when $\mu$ increases. The vertical dependence on ${\cal N}_{\star}$ mirrors
the increasing amount of removed escapers due to the longer $t_{\rm cc}$--times
for larger ${\cal N}_{\star}$.

Panels (c) and (d) deal with the individual masses of the escapers.  When one
heavy star escapes, it raises the average escaper mass. In the models $\mu
\gtrsim 10$, the {absolute} number of high--mass escapers is zero for most of
the runs.  This is the converse situation of panel (a): For massive stars, the
escape is difficult.  Just 1 or 2 heavy components appear and take away a
significant fraction of mass from the cluster. They are rather evaporated
objects than ejected in a close encounter, for their energies are relatively
small. Usually, the heavy escapers occur in the late post--collapse phase and
do not receive our attention here. From panel (d) and $\mu \leq 5$, we find
that $\approx$ 3\% of all escapers are heavy components which leave the system
before $t_{\rm cc}$.

Panel (e) presents the rates of escape within a crossing time for
different $\mu$.  The computation was adopted from Wielen (1975) as

\begin{eqnarray}
  \left \langle \frac{d{\cal N}_{\star}}{dt} \right \rangle
       = t_{\rm cr} \frac{\langle {\cal N}_{\rm esc}\rangle}
                         {\langle t_{\rm cc} \rangle} ,
\end{eqnarray}

where the brackets denote averages among the runs within a model and $t_{\rm
cr}$ is the average number of escapers per crossing time.  For equal--mass
clusters our rate is $\approx$ 0.2---0.3 in accordance with other $N$-body
results listed by \citet{GH94a}. The slight decline until $\mu \approx 3$ and
the stronger increase afterwards is in excellent agreement with the theoretical
expectation by \citet{Henon69} (Fig. 1). Though his models differ from ours in
$q$, the branch of the curve related to our models exhibits the same shape. We
interpret the curve such that two different mechanisms produce escapers in the
pre--collapse phase under consideration: Relaxation dominates if the individual
masses do not differ much, thus evaporation causes a steady mass loss from the
system.  When $\mu$ is increased slightly, the effect of evaporation cannot
advance so far. The escape rate is reduced then.  For higher $\mu$'s another
mechanism takes over: Massive stars exhibit a strong gravitational focusing.
The gravity of one single heavy particle attracts more light stars and its
energy is distributed among them.  So, a multitude of light stars easily gains
positive energy and heads for escape.  The frequent two--body encounters of one
heavy particle leads rather to ejections involving high energies than an
accumulation of small escape energies.

Panel (f) of Figure \ref{fig:escrate} shows the mean energy carried away by the
stars.  The near constancy suggests an independence on $\mu$, but a subdivision
into removed and potential escapers (not shown here) reveals differences
between the two groups.  For removed escapers, the mean energy increases as
much as a factor of 10 over the whole $\mu$--range.  This confirms the ejection
scenario explained for panel (e).  On the other hand, the potential escapers
show a constant but somewhat lower energies on average.  Since they make up the
larger fraction, the mean is depleted in the high--$\mu$ regime. 

Finally, we can conclude that models resembling the equal--mass model
tend to lose their mass by a slow evaporation process, while energetic
ejections outweight in high--$\mu$ models.  At $\mu \approx 3$, both
processes are exchanging their dominant role.  After mass segregation
has come to a stop near core bounce, three--body encounters in the
core start depleting the population of $\mh$--stars.

\section{Larger particle numbers} \label{ch:largen}

Some dependencies on different particle numbers have already been tacled in the
foregoing sections. Here we present a direct comparison of the cluster
evolution for the A--models first.  This model is close to the uniform mass
case and has the longest evolution time. Figure \ref{fig:phiforn} shows the raw
data of the minimum potential for five runs, i.e.\ we applied no smoothing or
averaging.

The most obvious feature, is the increasing core collapse time in pretty
accordance to the increasing relaxation time proportional to ${\cal
N}_{\star}/{\rm ln}(\gamma {\cal N}_{\star})$ (see section 8.1). Note that the
fluctuations of the data in the pre--collapse phase are smaller for higher
${\cal N}_{\star}$, because the global potential is smoother.  In the
post--collapse, the fluctuations are nearly the same, for the number of core
particles, $N_{\rm c}$, is of the same order for each of the five runs.

The assumed constancy of ${\cal N}_{\rm c}$ (where the subscript ``c'' stands
for core) leads to a second topic concerning the different amplitudes of the
potential minima. This has already been mentioned in connection with Figure
\ref{fig:tccmu2}, though the upper panel of Figure \ref{fig:fntcc} points to
the variations on ${\cal N}_{\star}$ in a more clear way. For models resembling
the equal--mass case, the maximum depth is a function of ${\cal N}_{\star}$.
The reason is that the collapse is only halted when the rate of energy
production in the core becomes the same as the energy rate going out via the
heat flux of the gravothermal instability \citep{Goodman87}. The outflowing
energy is produced by the formation of binaries in three--body encounters, and
this become important when the density is sufficiently high.  As the core
radius of a large--${\cal N}_{\star}$ cluster contains a larger number of stars
initially, it has to get rid of almost all of them. The final ${\cal N}_{\rm
c}$ at core bounce is nearly a constant and the evolution of the core radius
advances deeper in order to provide the density necessary for binary formation.
\citet{GH94b} described this scenario in terms of the fraction of core radius
to half--mass radius, $r_{\rm c}/r_{\rm h}$: the larger ${\cal N}_{\star}$, the
smaller that fraction. This is visible for models with $\mu$ close to unity in
Figure \ref{fig:fntcc}. Though, the lower panel suggests a slight increase of
core particles for large--$N$ models, the absolute values of ${\cal N}_{\rm c}$
stay in the order between 10 and 40.

\bfig
\C
\resizebox{\hsize}{!}{\includegraphics[bb=64 370 300 600,clip]
	{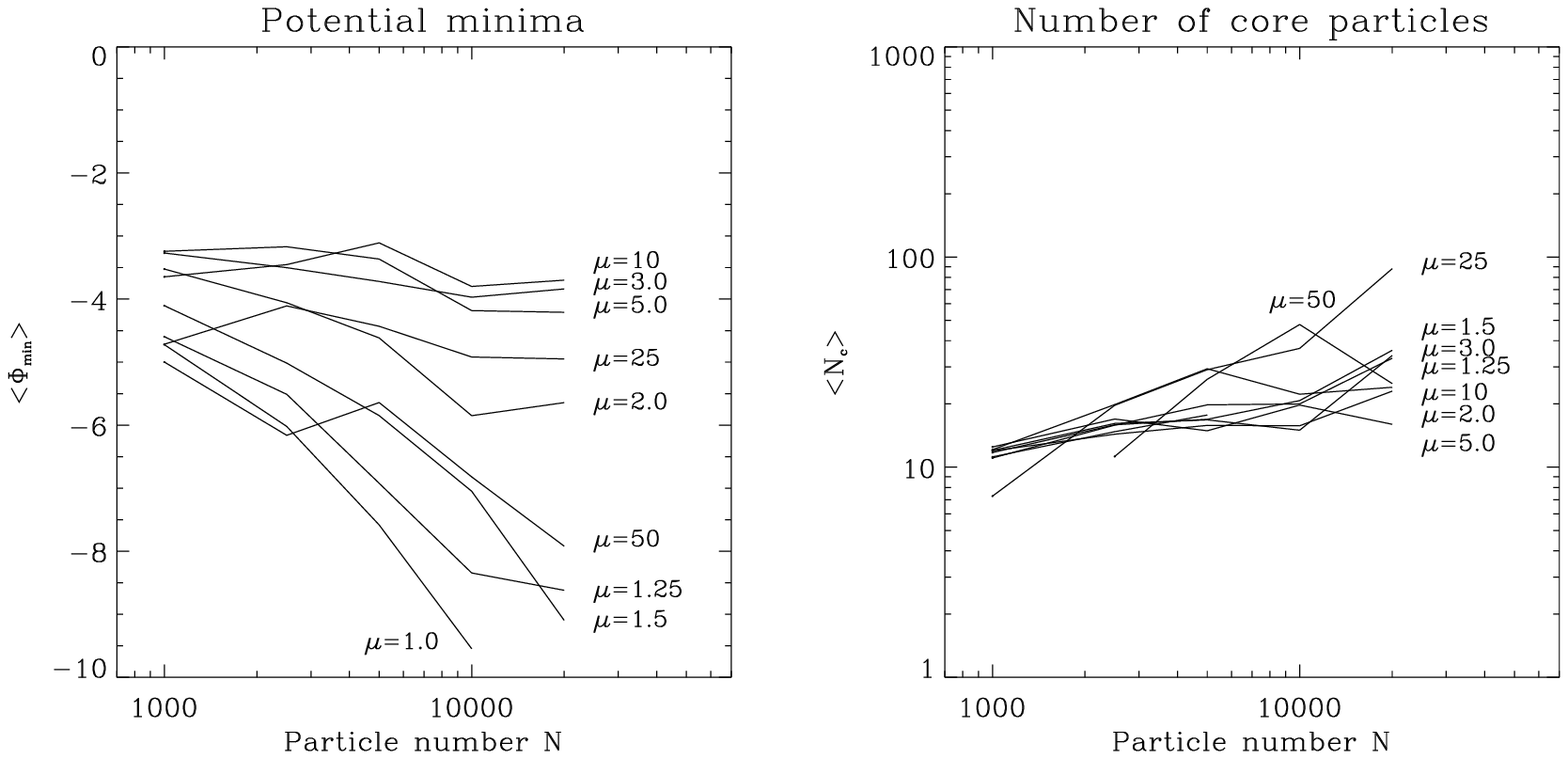}}
\resizebox{\hsize}{!}{\includegraphics[bb=319 370 549 602,clip]
	{grafiques/fntcc.ps}}
\caption{Upper panel: The mean potential minima at the moment
           of core bounce as as function of the total
           particle number ${\cal N}_{\star}$.
           Models close to the equal--mass case show
           a deeper peak when ${\cal N}_{\star}$ is rising than
           the high--$\mu$ models.
           Explanation is given in the text.
           Lower panel: The mean core number increases only
           slightly with ${\cal N}_{\star}$.
           Disregarding the two highest $\mu$, the
           number of minimum core particles is between
           10 and 40
\label{fig:fntcc} }
\EC
\efig

The situation looks somewhat different for models with high $\mu$, in
particular for $\mu$ = 25 and 50. The central potential shows its
deepest value when some few of the heavy masses have gathered in the
core, while the light stars have not changed their density
distribution. The presence of the massive stars causes the deeper
potential, and it appears not as profound as for nearly equal masses
(see Figure \ref{fig:tccmu2} for comparison). Therefore, the amplitude of the
collapse, $\langle \Phi_{\rm min}\rangle$, depends on ${\cal N}_{\rm
c}$ only weakly.

\subsection{The Coulomb logarithm}

Now, we will focus on the coefficient $\gamma$ in the Coulomb
logarithm, which is of relevant importance for the evolution of the
star cluster, for it is intimately connected with the time-scale
associated with the evolution of the cluster, the relaxation time
(Eq. 2-62 of \citealt{Spitzer87}). The value of this quantity, $\ln
\:(\gamma {\cal N}_{\star}$), has been estimated to be $\gamma = 0.11$
\citep{GH94a} for a single--mass cluster, but the variations on $\mu$
are not known precisely. One possible way to determine this quantity
is by comparing the evolution of the same model but for different
values of ${\cal N}_{\star}$.
                                                                  
\bfig
\C
\resizebox{\hsize}{!}{\includegraphics[bb=72 370 306 602,clip]
	{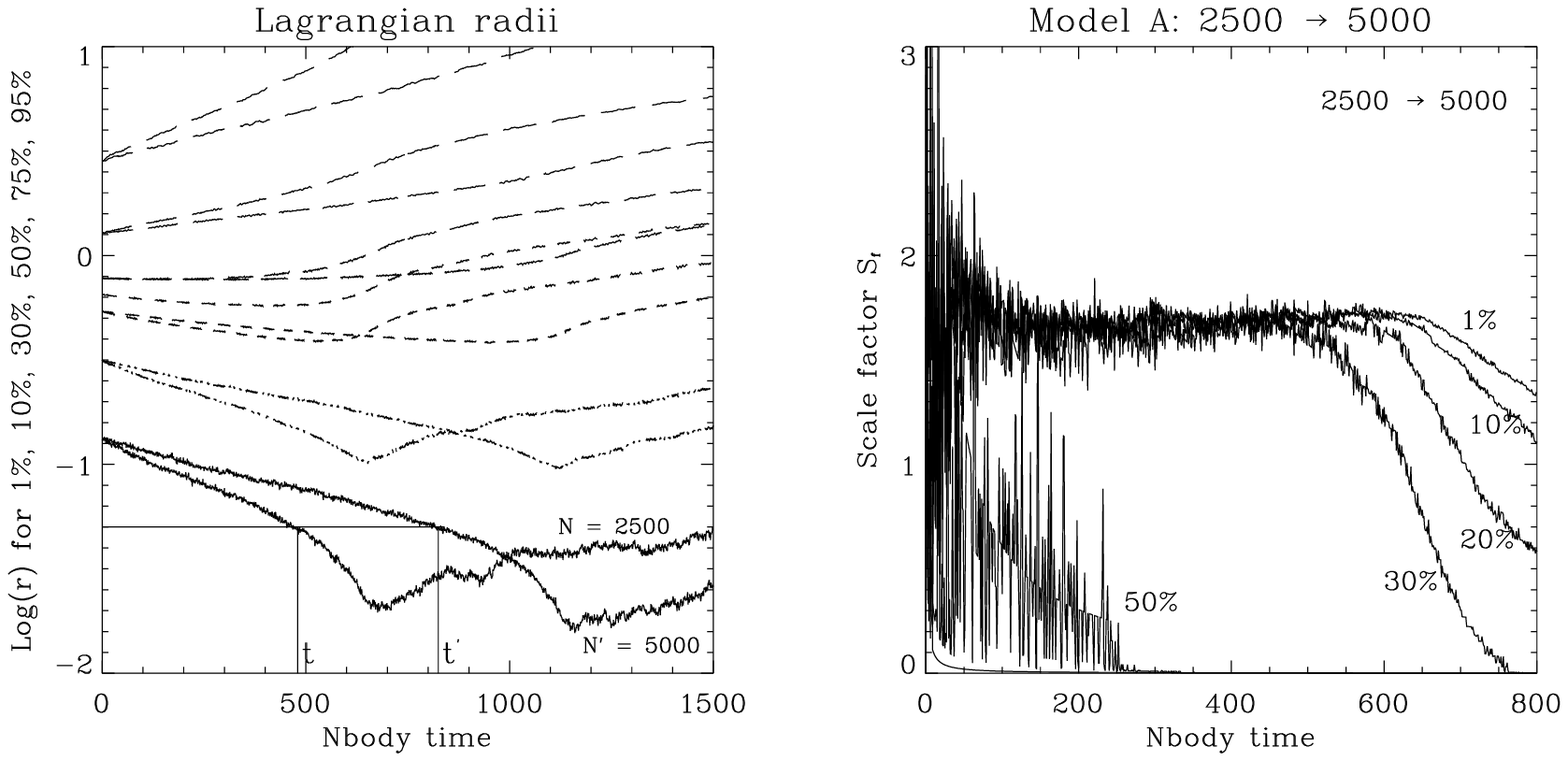}}\vspace{0.1cm}
\resizebox{\hsize}{!}{\includegraphics[bb=333 370 554 602,clip]
	{grafiques/fgamma2.ps}}
\caption{Upper panel: Five selected Lagrangian radii
           of Model A for ${\cal N}_{\star} = 2.5\cdot 10^3$ and
           ${\cal N}_{\star}^{\prime} = 5\cdot 10^3$. Lower panel:
           Scale factor $S_{\rm f}$ for this model computed from the
           comparison of the Lagrangian radii
\label{fig:gamma2} }
\EC
\efig

The upper panel of Figure \ref{fig:gamma2} shows the
ensemble--averaged Lagrangian radii of our Model A for two different
${\cal N}_{\star}$.  For the time $t$, the value of a Lagrangian
radius in the 2500--body model was determined, and then the
corresponding time $t^{\prime}$ at which the same value was reached in
the 5000--body model.  
 
The ratio of these two times is a scale factor, ${\cal S}_{\rm f}$,
that should be also equal to the ratio of the relaxation times for the
different ${\cal N}_{\star}$ before core rebounce:
\begin{eqnarray}
\frac{t^{\prime}}{t}  \equiv  {\cal S}_{\rm f}
                 := \frac{ {\cal N}_{\star}^{\prime}\, {\rm ln}(\gamma
                 {\cal N}_{\star}) } { {\cal N}_{\star}\, {\rm
                 ln}(\gamma {\cal N}_{\star}^{\prime}) }
                 . \label{eq:sf}
\end{eqnarray}
By repeting this procedure for each time step and for each Lagrangian
radius we get an ${\cal S}_{\rm f}$ that is plotted in the lower panel
of Figure \ref{fig:gamma2}. Before the core collapse, the time ratios
show a remarkable constancy (disregarding the initial settling
period), and the inner Lagrangian radii have a good agreement. The scale factor in the constant
pre--collapse range is ${\cal S}_{\rm f} = 1.675$ for this model.
With the definition $\nu := {\cal N}_{\star}^{\prime}/{\cal
N}_{\star}$, a re--arrangement of eq.~(\ref{eq:sf}) yields
\begin{eqnarray}
%
\gamma &=& \left( \frac{{\cal N}_{\star}}{({\cal N}_{\star}^{\prime})^{{\cal S}_{\rm f}/\nu}} \right) ^%
                     {{\nu}/({{\cal S}_{\rm f}-\nu})} .
                     \label{eq:gamma}
\end{eqnarray}

We computed the $\gamma$'s for each model using the data sets of
${\cal N}_{\star}=2.5 \cdot 10^3$ and ${\cal N}_{\star}^{\prime}=5
\cdot 10^3$. The results are given in Table \ref{tab.mass_seg4} 
and plotted in Figure \ref{fig:sfgamma}. The first two rows of the
Table define the $\mu$--model, the third gives the scale factor as
determined from the figures analog to
\ref{fig:gamma2}, the fourth is the error measured from the widths of
the fluctuating lines in the stable regime (horizontal part), the
fifth is the outer exponent in Eq.\,(\ref{eq:gamma}) with $\nu = 2$ and
the sixth are the resulting $\gamma$'s.

\bfig
\C
\resizebox{\hsize}{!}{\includegraphics[bb=75 370 300 600,clip]
	{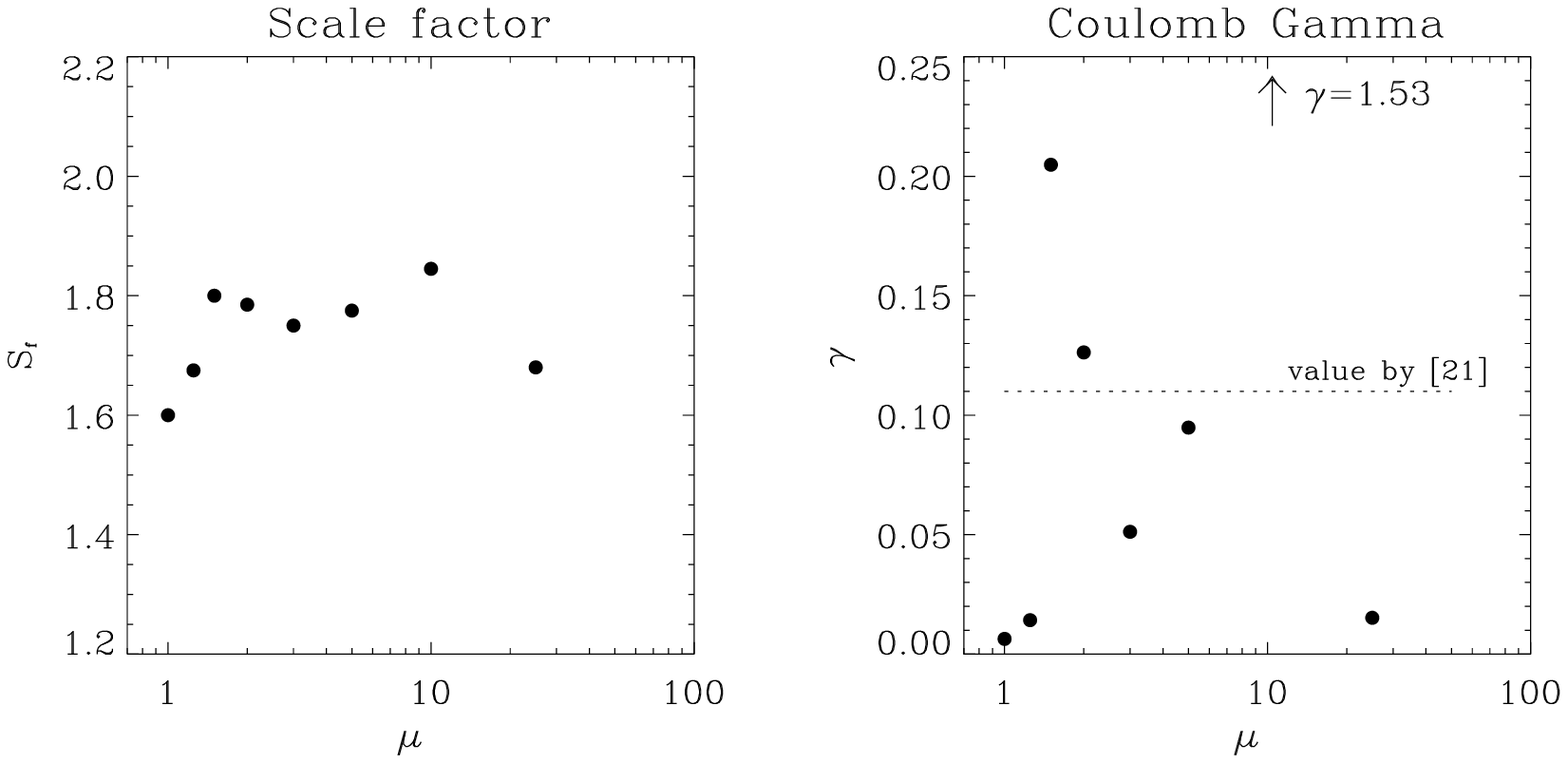}}
\resizebox{\hsize}{!}{\includegraphics[bb=326 370 554 600,clip]
	{grafiques/fsfgamma.ps}}
\caption{Upper panel: Scale factors measured from
           figures analog to \ref{fig:gamma2} for
           various $\mu$.
           Lower panel: Coulomb--$\gamma$ as determined
           from $N$-body simulations with different
           particle numbers from eq.~(\ref{eq:gamma}).
           The error bars are larger than the panel size
\label{fig:sfgamma} }
\EC
\efig

\begin{table*}
\begin{center}
\begin{tabular}{l|*{6}{c|}c}  \hline  \label{tab:gamma}
   model      &      A     &     B      &
                     C     &     D      &
                     E     &     F      &
                     G     \\
   $\mu$      &    1.25    &    1.5     &
                   2.0     &    3.0     &
                   5.0     &   10.0     &
                  25.0     \\ \hline
  ${\cal S}_{\rm f}$ &    1.675   &    1.800   &
                   1.785   &    1.750   &
                   1.775   &    1.845   &
                   1.680   \\
$\Delta {\cal S}_{\rm f}$& 0.045  &    0.010   &
                   0.035   &    0.030   &
                   0.175   &    0.045   &
                   0.190  \\
$\nu/({\cal S}_{\rm f}-\nu)$
             &  $-$6.154   &  $-$10.00  &
                $-$9.302   &  $-$8.000  &
                $-$8.889   & $-$12.903  &
                $-$6.250   \\
   $\gamma$  & {\bf 0.0142}&{\bf 0.2048}&
               {\bf 0.1263}&{\bf 0.0512}&
               {\bf 0.0948}&{\bf 1.5321}&
               {\bf 0.0152}\\
                              \hline
\end{tabular}
\end{center}
\caption{Scale factors and $\gamma$'s for the models
         with different $\mu$. The data were determined by comparing
         the data sets ${\cal N}_{\star} = 2500$ and ${\cal
         N}_{\star}^{\prime} = 5000$, giving a particle ratio $\nu=2$
\label{tab.mass_seg4}
}
\end{table*}

The comparison of the data sets ``$2.5\cdot 10^3 \to 5\cdot
10^3$'' provides a more accurate scale factor than the other data sets
with lower ${\cal N}_{\star}$, because they rest upon a higher
statistical significancy: Our set with ${\cal N}_{\star}=10^4$ can be
averaged over 4 runs only, and the $10^3$--set is biased to
low--${\cal N}_{\star}$ physics.

Very small variations in ${\cal S}_{\rm f}$ cause large fluctuations
in the two exponents of eq.~(\ref{eq:gamma}) and alter $\gamma$
significantly; the formula is very sensitive to both, ${\cal S}_{\rm
f}$ and $\nu$.  The error was computed as the difference of an
``upper'' and a ``lower'' $\gamma$ that results from the thickness of
the ${\cal S}_{\rm f}$--line.

As for different $\nu$, we present another example illustrating the
big variations when determining the Coulomb--$\gamma$. Table
\ref{tab.mass_seg5} shows the results for the equal--mass models using
various particle numbers ${\cal N}_{\star}$ and ${\cal
N}_{\star}^{\prime}$.  The ${\cal S}_{\rm f}$ shows a reasonable
behaviour for different $\nu$'s and is conform to the result by
\citet{GH94a} shown in the last column.  The resulting $\gamma$,
however, differs at least by a factor of 5.

The value $\gamma = 0.11$ mentioned above is actually found due to the
comparison of $N$-body and Fokker--Planck simulations of equal--mass
models (\citealt{GH94a}, Figure 6). From their $N$-body--$N$-body
simulations with particle numbers of ${\cal N}_{\star}=500$ and ${\cal
N}_{\star}^{\prime}=2000$ (their Figure 5), it is easy to extract an
${\cal S}_{\rm f} = 2.95$ and get a $\gamma = 0.098$. For the same
particle ratio, $\nu =4.0$, we obtain a $\gamma$ that is half of
theirs (last two columns in Table \ref{tab.mass_seg5}). But our absolute particle
numbers are 5 times larger. We can conclude from our analysis that the
choice of ${\cal N}_{\star}$ and ${\cal N}_{\star}^{\prime}$
essentially contributes to the Coulomb--$\gamma$. 

\begin{table*}
\begin{center}
\begin{tabular}{l|ccccc} 
Eq. Ms          & 1000 $\to$ 2500 & 1000 $\to$ 5000 &
                 2500 $\to$ 5000 & 2500 $\to 10,\!000$&
                \cite{GH94a} \\ \hline
    $\nu$      &    2.5    &    5.0    &    2.0    &    4.0  &    4.0   \\
    ${\cal S}_{\rm f}$&    1.96   &    3.21   &    1.60   &    3.09 &    2.95  \\
$\nu/({\cal S}_{\rm f}-\nu)$&$-$4.630&$-$2.793& $-$5.00   & $-$4.396& $-$3.810 \\
     $\gamma$  &    0.028  &    0.018  &    0.006  &   0.044 &    0.098 \\
    \hline
\end{tabular}
\end{center}
\caption{Comparison of our equal--mass models with four
         different particle numbers and the $N$-body--analysis
         by \citet{GH94a} in the last column
\label{tab.mass_seg5}
}
\end{table*}

Finally, we are left with a fair range of possible values for
$\gamma$. An estimate indicates that $\gamma$ ranges somewhere between
0.01 and 0.20.  A similar range was given by Giersz \& Heggie (1996)
for the case of a power--law mass spectrum: $0.016 \lesssim
\gamma \lesssim 0.26$.


\section{Variations of the mass fraction} \label{ch:results3}

So far we have presented results on mass segregation for various values of
$\mu$. This section deals with three additional series, in which we alter the
fraction of the heavy masses, $q$. Such kind of study has been discussed by
\citet{WJR00} with a Monte Carlo approach or by \citet{IW84}, who simulated
two--component clusters by means of Fokker--Planck modelling. They fixed $\mu$
to 2.0 and investigated the core collapse times as well as the achievement of
equipartition.  They showed that the evolution of the central potential in
units of the half--mass relaxation time, $t_{\rm rh}$, for 4 different $q$'s is
fastest when $q \approx 0.1$.

We present in this section the question whether equipartition can be achieved
between the two components while segregation is on work and a comparison of the
$N$-body data with the above mentioned literature.

We investigate the equipartition parameter
\begin{eqnarray}
  \xi = \frac{\mh\sigma_2^2}{\ml\sigma_1^2} ,  \label{eq:xi}
\end{eqnarray}
which gives the ratio of the kinetic energies between the heavy and
light stars in the core. At the start its value is about $\mu$ and
heads for unity. When both mass species find a state of energy
equipartition, $\xi = 1$ is reached, and we call the system
equipartition stable (after \citealt{Spitzer69}), otherwise a
$\xi_{\rm min}$ indicates the closest approach to it.

When examing a particular run, the data of $\xi$ is very noisy,
especially, for small--${\cal N}_{\star}$ simulations. The main cause
for this is the small number of particles inside the core radius. For
this reason, we decided to evaluate for the particles inside {twice}
the core radius, and then apply the smoothing procedure of section
\ref{ch:ensembles}.
We define $\xi_{\rm min} = 1$, if eq.~(\ref{eq:xi}) drops below unity
at any time during the evolution; otherwise we set it to the deepest
peak. Figure \ref{fig:equi_bf} shows an example for the parameter
$\xi$ from two models with ${\cal N}_{\star} = 104$. The
B--model does reach equipartition at $t = 1157$, so $\xi_{\rm min}$ is
set to 1; the F--model approaches to it down to $\xi_{\rm min} = 2.95$
at $t = 106$.

\bfig
\C
\resizebox{\hsize}{!}{\includegraphics[bb=78 370 304 600,clip]
	{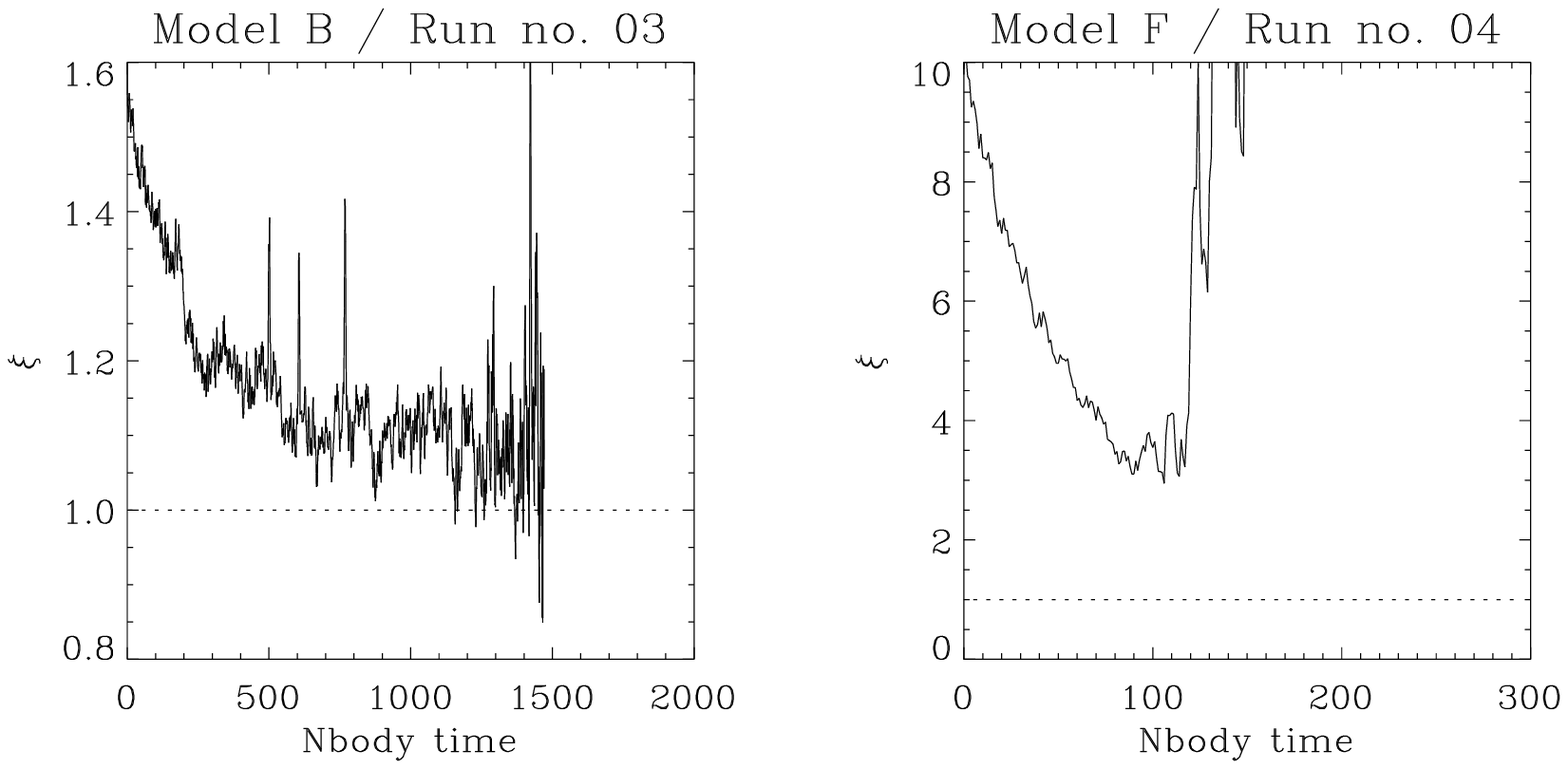}}
\resizebox{\hsize}{!}{\includegraphics[bb=334 370 553 600,clip]
	{grafiques/fequi_bf.ps}}
\caption{Equipartition parameter for two runs
           from model B (upper panel) and F (lower panel),
           respectively, containing $10,\!000$ particles. The data was
           smoothed out with a smooth width $w=5$ as described in
           Chapter \ref{ch:ensembles}
\label{fig:equi_bf} }
\EC
\efig

The average from all the runs' particular $\xi_{\rm min}$'s is denoted
by $\langle \xi_{\rm min} \rangle$ and taken as the most reliable data
for the model. The $\langle \xi_{\rm min}\rangle$'s are plotted
versus $\mu$ in Figure \ref{fig:ximu}. The symbols belong all to the
Series I with $q = 0.1$, but a different particle number. The solid
line connects the ${\cal N}_{\star}=2.5 \cdot 10^3$ values
(triangles). The additional Series III ($q=0.05$), IV ($q=0.2$), and V
($q=0.4$) are shown as solid lines. Error bars are given only for the
four sets with ${\cal N}_{\star}=2.5 \cdot 10^3$; they are smaller
than the symbols in most cases.

As expected, the graphic shows that equipartition takes place for
small $\mu$'s, but when $\mu$ becomes significantly greater than 2,
$\langle \xi_{\rm min} \rangle$ recedes from unity.  At $\mu=2$, about
half of the individual runs with $q=0.1$ did succeed to reach
$\xi_{\rm min} = 1$, at least for a moment.  Those runs, which did not
find a state of full equipartition, tried to reduce the kinetic
difference halfways to the core collapse, but then departed shortly
after the closest approach.

\begin{table}
\begin{center}
\begin{tabular}{lc|cc}
Series & $q$   &  $\mu_{\rm crit}$  \\  \hline
  III  & 0.05  &  2.49   &  \\
   I   & 0.10  &  2.03   &  $\left\{
                      \begin{array}{l}
                         2.032 \quad ({\cal N}_{\star} = 1000) \\
                         2.048 \quad ({\cal N}_{\star} = 2500) \\
                         2.197 \quad ({\cal N}_{\star} = 5000) \\
                         2.111 \quad ({\cal N}_{\star} = 10{\rm k})\\
                         1.762 \quad ({\cal N}_{\star} = 20{\rm k})
                      \end{array} \right.$   \\
  IV  & 0.20  &  1.87   &  \\
  V   & 0.40  &  1.75   &  \\
\label{tab.mass_seg6}
\end{tabular}
\end{center}
\caption{Values of $\mu$ at which $\xi_{\rm min} > 1.05$.
        Equipartition cannot be attained anymore. See text for further
        details}
\end{table}


\bFig
  \resizebox{12cm}{!}{%
          \includegraphics{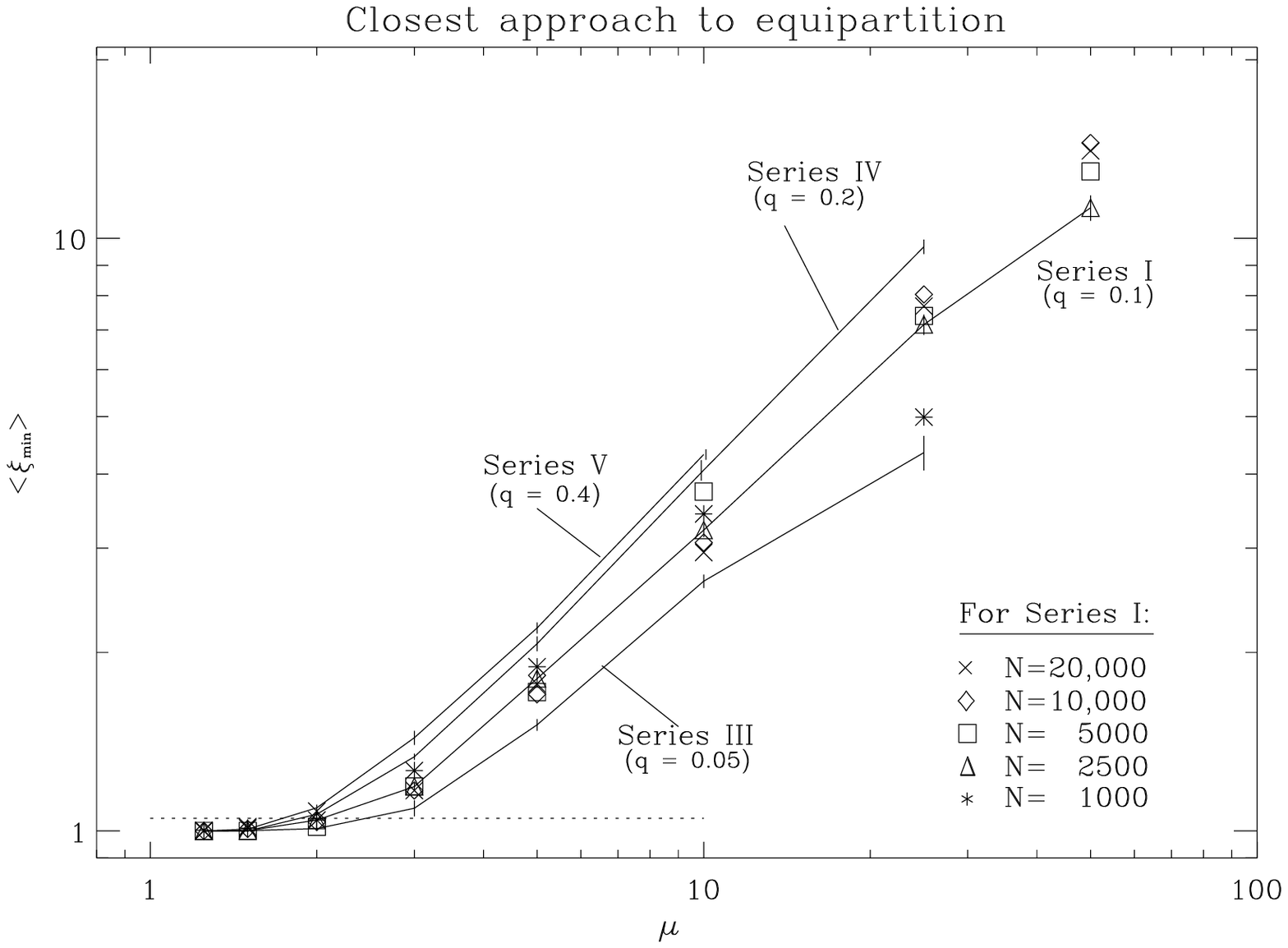}%
        }
  \hfill \parbox[b]{55mm}{%
\caption{The averaged minimum of the equipartition parameter
           $\xi$ for stars inside twice the core radius.
           When both mass components are in equipartition,
           $\xi_{\rm min}$ equals 1.
           The solid lines connect the data points for $q=0.4$ (Series
           V), $q=0.2$ (Series IV), $q=0.1$ (Series I with triangles),
           and $q=0.05$ (Series III).  These four simulation series
           were made with ${\cal N}_{\star}=2.5\cdot 10^3$ particles,
           and error bars are given for them.  The dotted line is an
           arbitrary threshold for equipartition stability at 1.05 as
           explained in the text}
\label{fig:ximu}
}
\eFig

That figure provides a good basis for the judgement on equipartition
stability. It is obvious that $\langle \xi_{\rm min}
\rangle$ varies for different fractions of heavy masses, $q$: The less
amount of heavy masses in a cluster, the closer equipartition is
reached. This is consistent with the results by both \citet{IW84} and
\citet{WJR00}. The latter explore an even wider range to very low
$q$'s down to 0.0015 (their set ``B'').

In order to check the stability border, we look now for a ``critical
$\mu$'' at which equipartition stability is \emph{not} given.

Instead of assuming that equipartition is happens for $\langle
\xi_{\rm min} \rangle = 1$, we will give a small tolerance for this and
define the point of ``turning away'' from equipartition at $\langle
\xi_{\rm min}\rangle = 1.05$, as \citet{WJR00} did.  It is denoted by
the dotted horizontal line in Figure \ref{fig:ximu}.  By linear
interpolation between the lower--next and upper--next data point, one
obtains that this threshold is exceeded at the points $\mu_{\rm crit}$
given in Table \ref{tab.mass_seg6}.

The value for Series I was averaged from all simulation sets for
different ${\cal N}_{\star}$.  All $\langle \xi_{\rm min}\rangle$'s
are plotted Figure \ref{fig:xiplane}, and fitted for a direct
comparison of the Monte Carlo results from \citet{WJR00}.  The
positions of our $\mu_{\rm crit}$ are marked by a small filled dot.
The apparent difference for the line by \citet{LF78} is just bacause
\citet{WJR00} defined their fraction of heavy masses as $\hat{q} =
\MH/\ML$, while we use $q=\MH/\MCL$ (see section \ref{ch:params}).
The graphics and the results are consistent with each other and ease
the comparison.

\bfig
\C
\resizebox{\hsize}{!}{\includegraphics
	{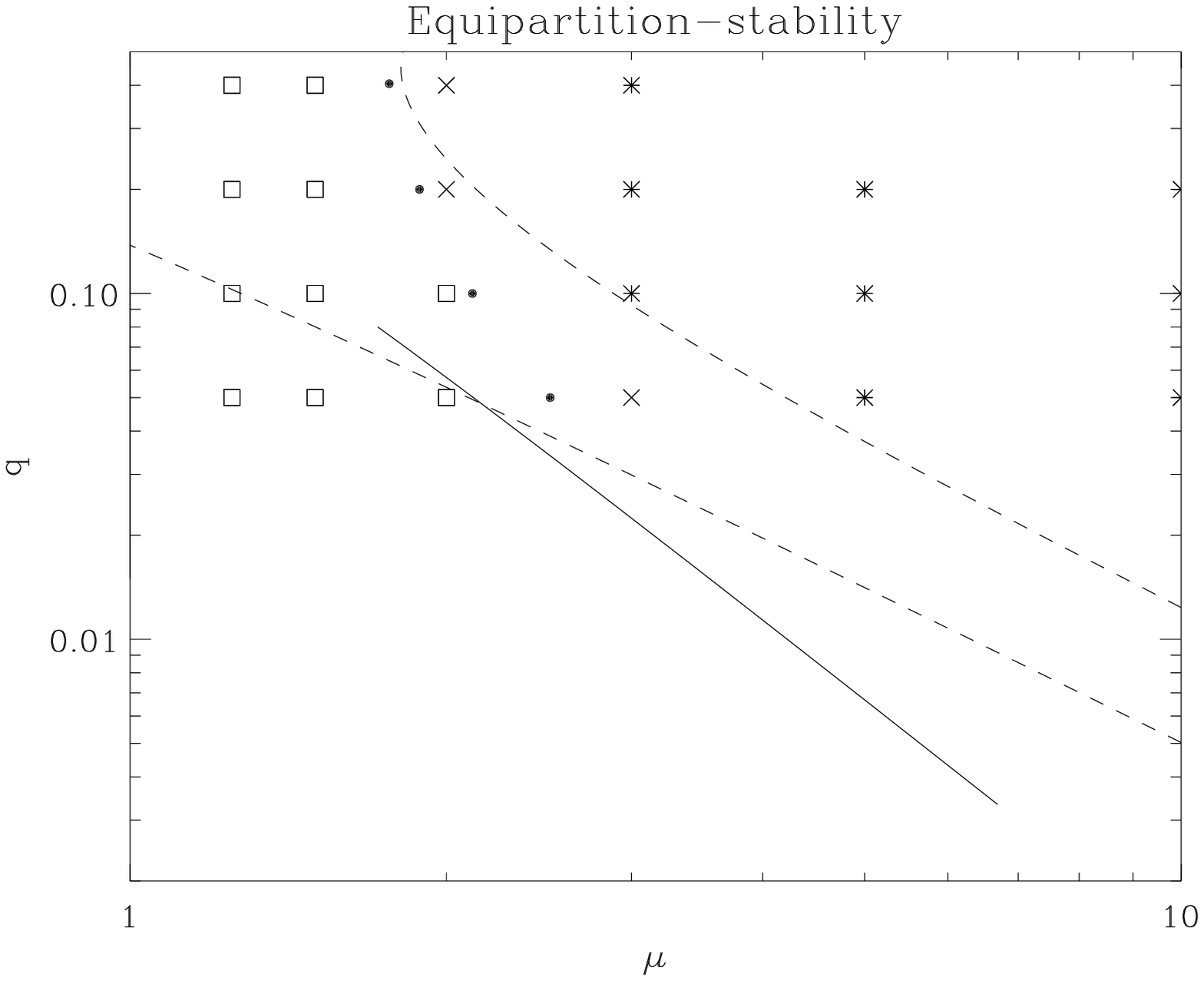}}
\caption{Closest approach to equipartition for the
           models in our parameter range of $\mu$ (mass ratio of
           heavy stars to light stars) and $q$ (mass in heavy
           stars relative to total cluster mass, see also Figure
           \ref{fig:phdmodels}). The figure has been adjusted
           (symbols, axes, and lines are accommodated as explained in
           the text) so that one can easily compare it with the
           results with those of \citet{WJR00} (Fig.~6).  The solid
           line is their stability criterion,
           Eq.~(\ref{eq.spitzer_stab_new}); the straight dashed line
           is from \citet{Spitzer69}, and the curved dashed line is
           from \citet{LF78}
\label{fig:xiplane} }
\EC
\efig

Although we have four points for $\mu_{\rm crit}$ in order to check
the equipartition boundary, they follow the theoretical function by
\citet{LF78} in fair agreement.

The formula suggested by \citet{WJR00} cannot be ruled out, for it is
based on models in a low--$q$ regime, which is difficult to access
with our $N$body simulations. However, our experimental data to show
that the criterion by \citet{Spitzer69} for the stability boundary,
appears too strong, especially at mass ratios $\mu$ close to 1.

$\mu_{\rm crit} \approx 2$ is the point which was already recognized
as the transition of an equipartition--dominated and a
gravothermal--dominated collapse (section \ref{ch:cctimes}).  Below
$\mu_{\rm crit}$, the core collapse proceeds slowlier than the theory
predicts, because equipartition governs the initial phase. When $\mu >
\mu_{\rm crit}$, the gravothermal instability always wins the
competition between the two effects: Before the thermal equilibrium
can fully be achieved, the massive stars have already segregated to
the centre and collapse independently from the light stars. With the
total fraction $q$ being high, the massive component appears almost
self--gravitating and is decoupled from the beginning.

\bfig
\C
\resizebox{\hsize}{!}{\includegraphics
	{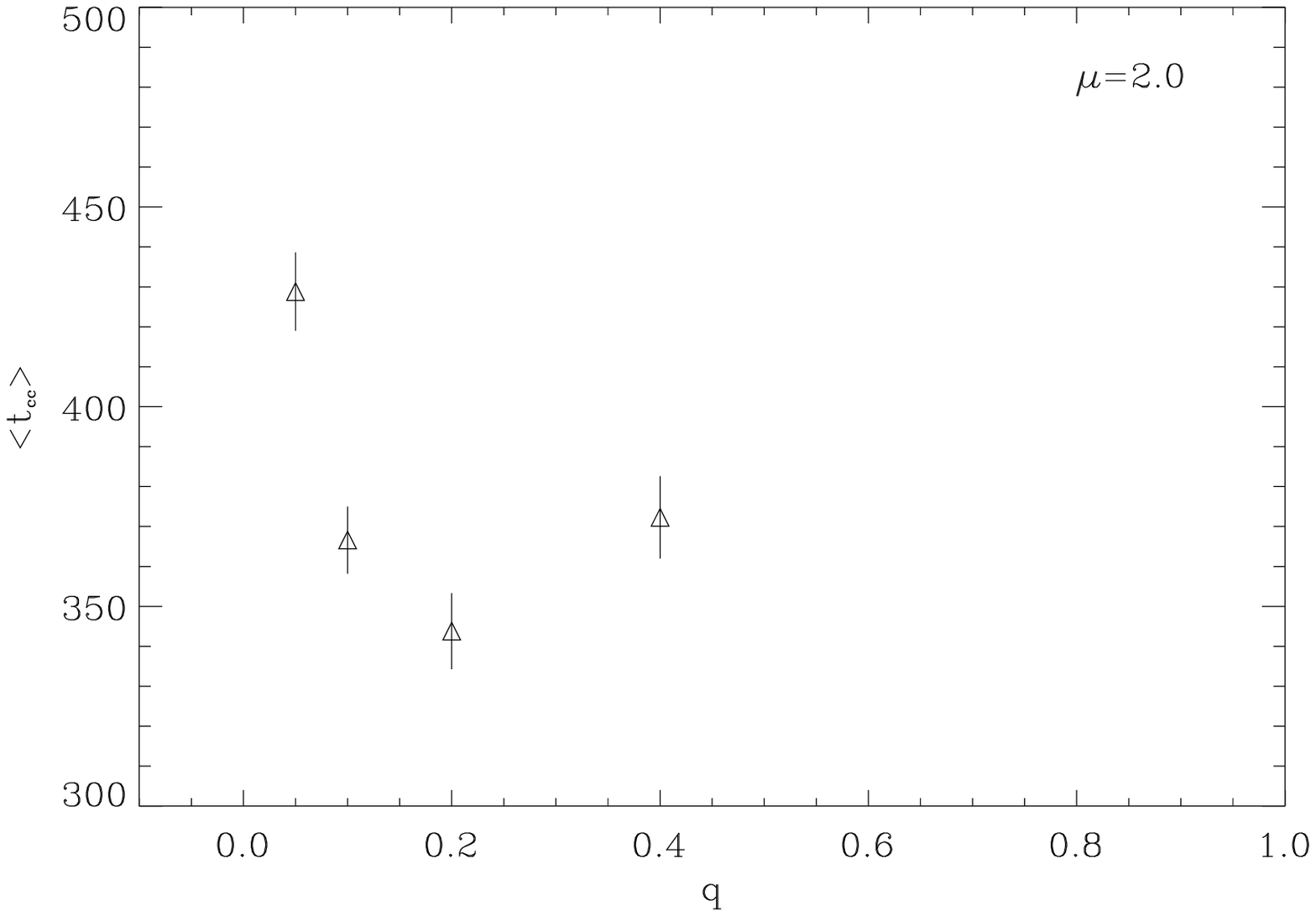}}
\caption{Core collapse time $t_{\rm cc}$ as a
           function of $q$. The individual mass ratio, $\mu$, was
           fixed to 2.0. The gas models by \citet{BI85} show a broad
           minimum for the core collapse times in terms of the
           relaxation time.  The value of $q$ for fastest evolution is
           $\approx$ 0.1--0.2. The results of the $N$-body models for
           the four series are discussed in the text. The time on the
           vertical axis is given in $N$-body units
\label{fig:tccq} }
\EC
\efig

\citet{BI85} showed that there is an ``optimal'' $q$ that favours the collapse
of a two--component system. If $q$ is small, the collapse of the heavier
component proceeds in the external field of the lighter stars, and the collapse
time decreases rapidly.  On the other hand, if $q$ is very large, the amount of
thermal energy contained in the core of the massive stars is so large that the
core of the light ones cannot receive all the redundant energy.  Hence the
tendency to collapse is postponed, because the heat gained has to be dumped
away into the halo and this process is slower than the tendency to
equipartition; see also \citep{Lee95}. Finally, the extreme cases $q \to 0$ and
$q \to 1$ approach the single--mass value for a negligible contribution of the
heavy and light component, respectively.  This is the reason why the core
collapse time attains a minimum at a certain fraction of the heavier stars
(Bettwieser \& Inagaki 1985). The $N$-body models confirm this (Figure
\ref{fig:tccq}).

\section{Conclusions}

We have studied mass segregation and equipartition in idealized star
clusters containing two mass components. The results provide potential
insight into the evolution of young star clusters, dynamics of star
forming regions, and the degree of initial segregation which is needed
to match the observations.  Also, this study is of theoretical
interest to check the classical ideas on equipartition and the
thermodynamic behaviour of self--gravitating systems.

The parameter space was widely analysed for both parameters using Plummer
models: the ratio of individual star masses, $\mu = \mh/\ml$, that was varied
from 1.25 to 50; and the fraction of the total heavy masses, $q = \MH/\MCL$,
that was chosen to be 0.05, 0.1, 0.2, or 0.4. We also directed our attention to
the value $q=0.1$, which was proved by \citet{IW84} to exhibit the fastest
evolution time.  For this fraction, we modelled different particle numbers,
with ${\cal N}_{\star}$ ranging from $10^3$ to $2 \cdot 10^4$.  For all other
cases we used a particle number of ${\cal N}_{\star}=2.5\cdot 10^3$.  The
statistical significancy of the data was considerably improved by
ensemble--averages: A large number of physically equivalent runs differing just
by a random seed for an initial setup was performed, and the data base was
gathered for an overall average.

For the first time we have empirically checked the borders between an
equipartition stable and unstable configuration \citep{Spitzer69} with the
accurate direct $N$-body method.  His criterion involves the parameters $q$ and
$\mu$ and is given in Eq.~(\ref{eq.spitzer_stab}). Moreover, we determined the
ranges of the longest and shortest core collapse time by comparing different
initial setups of the particles. 

From the results of our simulations we draw the following conclusions
on the global cluster evolution:
\begin{enumerate}
\item The evolution of a self--gravitating system
with two mass--components depends strongly on the ratio of the
individual masses, $\mu = \mh/\ml$.  For $q=0.1$, when $\mu$ is larger than
$\approx$ 2, the heavy masses fall to the centre and reduce the
relaxation time proportional to $1/\mu$.  Subsidiary, the core
collapse time is shortened in the same way (Fig.~\ref{fig:tccmu}).
Smaller $\mu$'s go into equipartition which slows down the evolution.
Large $\mu$'s give rise to a small number of heavy
particles, and the situation turns into a process of dynamical
friction.
\item There is a competition between equipartition
of energies and gravothermal instability.  If $\mu$ is close to unity,
equipartition hinders the gravothermal collapse in the initial phase,
but will never prevent it.  As the heavy stars congregate in the
central regions, they decouple from the light component and perform
their own collapse.  When $\mu$ exceeds a critical value of about $2$,
equipartition can never be achieved (Fig.~\ref{fig:ximu}).
\item The boundary between the stable and unstable
regimes is close to the theory by \citet{LF78} when $q$ is $\gtrsim$
0.2.  For $q$ lower than $\approx 0.1$, other works on
equipartition \citep{WJR00} seem to show a better agreement, but it is
difficult to assess their range with $N$-body simulations because the
absolute number of very heavy particles is too low in our standard
${\cal N}_{\star}$ (Fig.~\ref{fig:xiplane}).  When $q$ becomes larger
than 0.1, the criterion given by \citet{Spitzer69} appears too strong,
and his theory fails: According to Eq.~(\ref{eq.spitzer_stab}) no
equipartition should be possible at all, while our models as well as
the Monte Carlo simulations by
\citet{WJR00} show the contrary.  The cause is probably that Spitzer
assumed a {global} equilibrium when deriving the analytical
formul{\ae}. His basic ideas on the processes, however, are still in fair
accordance with our results.
\item Segregation proceeds on the relaxation time
scale simultaneously with the evolution of the cluster. The maximum
level of segregation is attained when the core has collapsed and
begins to expand. Then, the mass shells in the cluster have also
adjusted to a stable balance of an in--going and outgoing mass flux
(Fig.\ref{fig:fallshell}). After the core collapse, the degree of
stratification remains constant.

\item The escape rate shows two branches along
the $\mu$--axis indicating two different mechanisms being on work
(Fig.~\ref{fig:escrate}e).  For mass ratios resembling the equal--mass
case, the escape seems to be governed by evaporational effects in the
pre--collapse phase.  Since the core collapse time decreases for a
moderately rising $\mu$, evaporation does not advance far, such that
the mass loss is also reduced. For high $\mu$'s ($> 3$), we believe
that escapes are rather a matter of ejections; the very massive
particles distribute their kinetic energy to a large number of stars,
and the escape rate increases roughly with $\mu$. The details of
these processes, however, need more analysis tracing the path of the
escapers or a countercheck with other simulation methods.  As for the
equal--mass case, the escape rate is consistent with
other $N$-body simulations \citep{Wielen75,GH94a,BaumgardtEtAl02}.
\item Light masses play an important role in the heat
transfer from the core to the halo. Even a small fraction of them
moderates the heat flux between the central source and the outer sink.
The ``optimal'' fraction $q$ of heavy to light stars is about
0.15--0.20 in accord with previous results from gaseous models
(Fig.~\ref{fig:tccq}).
%
\end{enumerate}

Our results gained with the $N$-body method confirm previous
simulations with other techniques (gas models, Monte Carlo), but also
reveal hidden effects like the small ``deceleration'' of the
gravothermal collapse due to equipartition.  Further astrophysical
assumptions are likely to alter the results, e.g.\ the introduction of
primordial binaries in various fractions, a tidal field that
accelerates the collapse as well as the dissolution of clusters, or
rotation. Multiple mass components or a continuous spectrum make the
analysis more complex but are important for the understanding of the
observations. Our variation of $\mu$ can principally be generalized by
merging into a multi--mass model, but care has to be taken when
defining the parameters. The new parameters appearing for the models
are to be checked, e.g.\ the slope of the initial mass function (if a
power--law is assumed), and how the ratio of the highest mass to the
mean mass, $m_{\rm max}/ \langle m\rangle$, influences the
segregation.

\section*{Acknowledgements}

This work has been supported by Sonderforschungsbereich (SFB) 439 (subproject
A5) of German Science Foundation (DFG) at the University of Heidelberg. The
computations here were carried out at the CRAY T3E parallel computer at the
Hochleistungsrechenzentrum in Stuttgart\footnote{\tt http://www.hlrs.de/}. 

The work of PAS has been supported in the framework of the Third Level
Agreement between the DFG (Deutsche Forschungsgemeinschaft) and the IAC
(Instituto de Astrof\'\i sica de Canarias). PAS would like to thank Marc
Freitag for both fruitful discussions and very useful comments for the
elaboration and redaction of the text of the work, settled during his visit in
February 2006 to the Institute of Astronomy in Cambridge.  This was possible on
account of an invitation by Marc Freitag; thanks -again- Marc.

\appendix

\section{Appendix: Data tables}

The following tables contain the mean values of the
simulations of star clusters and are organised as
follows. The title specifies the Series, $q$,
and particle number.\\
Column 1: model name and $\mu$\\
Column 2: $\langle t_{\rm cc}\rangle$, mean core collapse time\\
Column 3: $\langle \Phi_{\rm min} \rangle$,
                    depth of central potential\\
Column 4: $\langle r_{\rm c} \rangle$,
                    mean core radius in the moment of core bounce\\
Column 5: $\langle {\cal N}_{\rm c} \rangle$,
                    number of particles in the core at the time of
                    smallest core radius\\
Column 6: $\langle \xi_{\rm min} \rangle$,
                    mean of the closest approach to equipartition\\
Column 7: $\langle {\cal N}_{\rm esc} \rangle$,
                    number of escaped stars until core collapse time\\
Column 8: $\langle E_{\rm esc} \rangle$, mean energy of escapers\\


\begin{table*}
\begin{center} 
\begin{tabular}{l|*{6}{r@{ $\pm$}r|}r@{ $\pm$}r}
\multicolumn{10}{l}{\textbf{Equal--mass: ${\cal N}_{\star}$ = 1000}} \\ \hline
model         & \multicolumn{2}{c|}{$\langle t_{\rm cc}\rangle $}
              & \multicolumn{2}{c|}{$\langle \Phi_{\rm min} \rangle $}
              & \multicolumn{2}{c|}{$\langle r_{\rm c} \rangle $}
              & \multicolumn{2}{c|}{$\langle {\cal N}_{\rm c} \rangle $}
              & \multicolumn{2}{c|}{$\langle \xi_{\rm min} \rangle $}
              & \multicolumn{2}{c|}{$\langle {\cal N}_{\rm esc} \rangle $}
              & \multicolumn{2}{c}{$\langle E_{\rm esc} \rangle $} \\ \hline
 EQMASS01     & {\bf 345.8}  &  5.2   &      
                {\bf $-$4.72}&  0.10  &      
                {\bf 0.0208} & 0.0020 &      
                {\bf 11.2}   &  0.4   &      
                \multicolumn{2}{c|}{---}&    
                {\bf 26.6}   &  1.1   &      
                {\bf 0.88}   & 0.18   \\     
                                         \hline
\end{tabular}
\end{center}

\end{table*}


\begin{table*}
\begin{center} 
\begin{tabular}{l|*{6}{r@{ $\pm$}r|}r@{ $\pm$}r}
\multicolumn{10}{l}{\textbf{Equal--mass: ${\cal N}_{\star}$ = 2500}} \\ \hline
model         & \multicolumn{2}{c|}{$\langle t_{\rm cc}\rangle $}
              & \multicolumn{2}{c|}{$\langle \Phi_{\rm min} \rangle $}
              & \multicolumn{2}{c|}{$\langle r_{\rm c} \rangle $}
              & \multicolumn{2}{c|}{$\langle {\cal N}_{\rm c} \rangle $}
              & \multicolumn{2}{c|}{$\langle \xi_{\rm min} \rangle $}
              & \multicolumn{2}{c|}{$\langle {\cal N}_{\rm esc} \rangle $}
              & \multicolumn{2}{c}{$\langle E_{\rm esc} \rangle $} \\ \hline
 EQMASS02     & {\bf 716.0}  & 12.0   &      
                {\bf $-$6.02}&  0.16  &      
                {\bf 0.0086} & 0.0006 &      
                {\bf 14.8}   &  0.8   &      
                \multicolumn{2}{c|}{---}&    
                {\bf 64.2}   &  3.1   &      
                {\bf 0.89}   & 0.24   \\     
                                         \hline
\end{tabular}
\end{center}
\end{table*}


\begin{table*}
\begin{center}
\begin{tabular}{l|*{6}{r@{ $\pm$}r|}r@{ $\pm$}r}
\multicolumn{10}{l}{\textbf{Equal--mass: ${\cal N}_{\star}$ = 5000}} \\ \hline
model         & \multicolumn{2}{c|}{$\langle t_{\rm cc}\rangle $}
              & \multicolumn{2}{c|}{$\langle \Phi_{\rm min} \rangle $}
              & \multicolumn{2}{c|}{$\langle r_{\rm c} \rangle $}
              & \multicolumn{2}{c|}{$\langle {\cal N}_{\rm c} \rangle $}
              & \multicolumn{2}{c|}{$\langle \xi_{\rm min} \rangle $}
              & \multicolumn{2}{c|}{$\langle {\cal N}_{\rm esc} \rangle $}
              & \multicolumn{2}{c}{$\langle E_{\rm esc} \rangle $} \\ \hline
 EQMASS05     & {\bf 1210.2} & 10.6   &      
                {\bf $-$7.58}&  0.27  &      
                {\bf 0.0045} & 0.0003 &      
                {\bf 17.7}   &  0.8   &      
                \multicolumn{2}{c|}{---}&    
                {\bf 117.1}  &  5.0   &      
                {\bf  0.71}  &  0.26  \\     
                                         \hline
\end{tabular}
\end{center}

\end{table*}


\begin{table*}
\begin{center}
\begin{tabular}{l|*{6}{r@{ $\pm$}r|}r@{ $\pm$}r}
\multicolumn{10}{l}{\textbf{Equal--mass: ${\cal N}_{\star}$ = 10,000}} \\ \hline
model         & \multicolumn{2}{c|}{$\langle t_{\rm cc}\rangle $}
              & \multicolumn{2}{c|}{$\langle \Phi_{\rm min} \rangle $}
              & \multicolumn{2}{c|}{$\langle r_{\rm c} \rangle $}
              & \multicolumn{2}{c|}{$\langle {\cal N}_{\rm c} \rangle $}
              & \multicolumn{2}{c|}{$\langle \xi_{\rm min} \rangle $}
              & \multicolumn{2}{c|}{$\langle {\cal N}_{\rm esc} \rangle $}
              & \multicolumn{2}{c}{$\langle E_{\rm esc} \rangle $} \\ \hline
 EQMASS10     & {\bf 2312.9} & 34.7   &      
                {\bf $-$8.75}&  0.47  &      
                {\bf 0.0025} & 0.0004 &      
                {\bf 21.3}   &  2.2   &      
                \multicolumn{2}{c|}{---}&    
                {\bf 256.3}  &  7.1   &      
                {\bf  2.55}  &  1.43  \\     
                                         \hline
\end{tabular}
\end{center}

\end{table*}


%
\begin{table*}
\begin{center}
\begin{tabular}{l|*{6}{r@{ $\pm$}r|}r@{ $\pm$}r}
\multicolumn{10}{l}{\textbf{Series I: $q$ = 0.1, ${\cal N}_{\star}$ = 1000}} \\ \hline
model ($\mu$) & \multicolumn{2}{c|}{$\langle t_{\rm cc}\rangle $}
              & \multicolumn{2}{c|}{$\langle \Phi_{\rm min} \rangle $}
              & \multicolumn{2}{c|}{$\langle r_{\rm c} \rangle $}
              & \multicolumn{2}{c|}{$\langle {\cal N}_{\rm c} \rangle $}
              & \multicolumn{2}{c|}{$\langle \xi_{\rm min} \rangle $}
              & \multicolumn{2}{c|}{$\langle {\cal N}_{\rm esc} \rangle $}
              & \multicolumn{2}{c}{$\langle E_{\rm esc} \rangle $} \\ \hline
 A (1.25)     & {\bf 309.1}  &  5.8   &      
                {\bf $-$4.60}&  0.11  &      
                {\bf 0.0206} & 0.0009 &      
                {\bf 11.1}   &  0.4   &      
                {\bf  1.002} &  0.002 &      
                {\bf 22.9}   &  1.2   &      
                {\bf 0.70}   & 0.17   \\     
 B (1.5)      & {\bf 256.0}  &  5.1   &
                {\bf $-$4.10}&  0.07  &
                {\bf 0.0275} & 0.0011 &
                {\bf 11.7}   &  0.3   &
                {\bf  1.011} &  0.005 &
                {\bf 18.4}   &  1.1   &
                {\bf 0.70}   & 0.11   \\
 C (2.0)      & {\bf 162.6}  &  3.5   &
                {\bf $-$3.52}&  0.04  &
                {\bf 0.0400} & 0.0016 &
                {\bf 12.0}   &  0.5   &
                {\bf  1.043} &  0.010 &
                {\bf  9.5}   &  0.5   &
                {\bf 0.78}   & 0.24   \\
 D (3.0)      & {\bf  86.1}  &  2.1   &
                {\bf $-$3.27}&  0.06  &
                {\bf 0.0596} & 0.0026 &
                {\bf 12.0}   &  0.6   &
                {\bf  1.264} &  0.033 &
                {\bf  4.8}   & 0.4    &
                {\bf 1.02}   & 0.26   \\
 E (5.0)      & {\bf  45.8}  &  1.4   &
                {\bf $-$3.24}&  0.06  &
                {\bf 0.0797} & 0.0040 &
                {\bf 12.5}   &  0.8   &
                {\bf  1.892} &  0.071 &
                {\bf  0.0}   &  0.0   &
                {\bf 0.00}   & 0.00   \\
 F (10.0)     & {\bf  25.4}  &  0.9   &
                {\bf $-$3.65}&  0.08  &
                {\bf 0.0954} & 0.0039 &
                {\bf 12.1}   &  0.7   &
                {\bf  3.426} &  0.187 &
                {\bf  4.1}   &  0.4   &
                {\bf 1.28}   & 0.33   \\
 G (25.0)     & {\bf  18.0}  &  0.7   &
                {\bf $-$4.72}&  0.12  &
                {\bf 0.0785} & 0.0030 &
                {\bf  7.3}   &  0.4   &
                {\bf  4.992} &  0.237 &
                {\bf 12.9}   &  0.2   &
                {\bf 1.09}   & 0.12   \\ \hline
\end{tabular}
\end{center}
\end{table*}


%
\begin{table*}
\begin{center}
\begin{tabular}{l|*{6}{r@{ $\pm$}r|}r@{ $\pm$}r}
\multicolumn{10}{l}{\textbf{Series I: $q$ = 0.1, ${\cal N}_{\star}$ = 2500}} \\ \hline
model ($\mu$) & \multicolumn{2}{c|}{$\langle t_{\rm cc}\rangle $}
              & \multicolumn{2}{c|}{$\langle \Phi_{\rm min} \rangle $}
              & \multicolumn{2}{c|}{$\langle r_{\rm c} \rangle $}
              & \multicolumn{2}{c|}{$\langle {\cal N}_{\rm c} \rangle $}
              & \multicolumn{2}{c|}{$\langle \xi_{\rm min} \rangle $}
              & \multicolumn{2}{c|}{$\langle {\cal N}_{\rm esc} \rangle $}
              & \multicolumn{2}{c}{$\langle E_{\rm esc} \rangle $} \\ \hline
 A (1.25)     & {\bf 677.8}  &  8.1   &      
                {\bf $-$5.51}&  0.12  &      
                {\bf 0.0100} & 0.0005 &      
                {\bf 15.9}   &  0.7   &      
                {\bf  1.000} &  0.000 &      
                {\bf 61.3}   &  2.0   &      
                {\bf 0.77}   & 0.27   \\     
 B (1.5)      & {\bf 558.4}  &  8.8   &
                {\bf $-$5.02}&  0.11  &
                {\bf 0.0137} & 0.0006 &
                {\bf 15.9}   &  0.8   &
                {\bf  1.001} &  0.000 &
                {\bf 45.2}   &  2.2   &
                {\bf 0.55}   & 0.14   \\
 C (2.0)      & {\bf 366.6}  &  8.4   &
                {\bf $-$4.06}&  0.11  &
                {\bf 0.0207} & 0.0011 &
                {\bf 14.4}   &  1.0   &
                {\bf  1.043} &  0.028 &
                {\bf 24.9}   &  1.6   &
                {\bf 1.18}   & 0.28   \\
 D (3.0)      & {\bf 180.1}  &  2.9   &
                {\bf $-$3.50}&  0.10  &
                {\bf 0.0336} & 0.0019 &
                {\bf 16.2}   &  2.3   &
                {\bf  1.188} &  0.017 &
                {\bf  9.7}   & 0.6    &
                {\bf 0.70}   & 0.12   \\
 E (5.0)      & {\bf  86.1}  &  3.3   &
                {\bf $-$3.17}&  0.10  &
                {\bf 0.0567} & 0.0043 &
                {\bf 17.0}   &  1.4   &
                {\bf  1.807} &  0.056 &
                {\bf  7.3}   &  0.8   &
                {\bf 1.43}   & 0.27   \\
 F (10.0)     & {\bf  40.2}  &  2.0   &
                {\bf $-$3.45}&  0.15  &
                {\bf 0.0811} & 0.0054 &
                {\bf 19.9}   &  1.8   &
                {\bf  3.218} &  0.086 &
                {\bf  8.0}   &  0.7   &
                {\bf 0.89}   & 0.12   \\
 G (25.0)     & {\bf  21.8}  &  1.1   &
                {\bf $-$4.11}&  0.18  &
                {\bf 0.0905} & 0.0048 &
                {\bf 19.7}   &  1.7   &
                {\bf  7.150} &  0.282 &
                {\bf 12.4}   &  0.9   &
                {\bf 0.86}   & 0.15   \\
 H (50.0)     & {\bf  17.9}  &  1.1   &
                {\bf $-$6.16}&  0.36  &
                {\bf 0.0792} & 0.0039 &
                {\bf 11.7}   &  1.0   &
                {\bf 11.245} &  0.556 &
                {\bf 25.2}   &  2.6   &
                {\bf 0.98}   & 0.15   \\ \hline
\end{tabular}
\end{center}
\end{table*}

%

%
\begin{table*}
\begin{center}
\begin{tabular}{l|*{6}{r@{ $\pm$}r|}r@{ $\pm$}r}
\multicolumn{10}{l}{\textbf{Series I: $q$ = 0.1, ${\cal N}_{\star}$ = 5000}} \\ \hline
model ($\mu$) & \multicolumn{2}{c|}{$\langle t_{\rm cc}\rangle $}
              & \multicolumn{2}{c|}{$\langle \Phi_{\rm min} \rangle $}
              & \multicolumn{2}{c|}{$\langle r_{\rm c} \rangle $}
              & \multicolumn{2}{c|}{$\langle {\cal N}_{\rm c} \rangle $}
              & \multicolumn{2}{c|}{$\langle \xi_{\rm min} \rangle $}
              & \multicolumn{2}{c|}{$\langle {\cal N}_{\rm esc} \rangle $}
              & \multicolumn{2}{c}{$\langle E_{\rm esc} \rangle $} \\ \hline
 A (1.25)     & {\bf 1183.7} & 17.7   &      
                {\bf $-$6.92}&  0.18  &      
                {\bf 0.0055} & 0.0004 &      
                {\bf 19.8}   &  0.7   &      
                {\bf  1.000} &  0.000 &      
                {\bf 103.9}  &  5.0   &      
                {\bf 0.44}   & 0.13   \\     
 B (1.5)      & {\bf 993.8}  & 10.3   &
                {\bf $-$5.85}&  0.12  &
                {\bf 0.0071} & 0.0003 &
                {\bf 16.9}   &  1.0   &
                {\bf  1.000} &  0.000 &
                {\bf 76.9}   &  3.2   &
                {\bf 0.57}   & 0.14   \\
 C (2.0)      & {\bf 642.6}  &  4.4   &
                {\bf $-$4.62}&  0.20  &
                {\bf 0.0117} & 0.0011 &
                {\bf 15.8}   &  1.4   &
                {\bf  1.016} &  0.008 &
                {\bf 38.0}   &  1.9   &
                {\bf 0.47}   & 0.09   \\
 D (3.0)      & {\bf 331.5}  & 11.6   &
                {\bf $-$3.72}&  0.08  &
                {\bf 0.0204} & 0.0013 &
                {\bf 16.8}   &  1.4   &
                {\bf  1.189} &  0.037 &
                {\bf 19.3}   & 2.6    &
                {\bf 1.08}   & 0.45   \\
 E (5.0)      & {\bf 157.9}  &  6.3   &
                {\bf $-$3.36}&  0.12  &
                {\bf 0.0302} & 0.0025 &
                {\bf 14.9}   &  1.1   &
                {\bf  1.715} &  0.039 &
                {\bf 14.2}   &  1.9   &
                {\bf  1.38}  &  0.27  \\
 F (10.0)     & {\bf  64.3}  &  3.5   &
                {\bf $-$3.11}&  0.09  &
                {\bf 0.0742} & 0.0055 &
                {\bf 29.4}   &  2.8   &
                {\bf  3.738} &  0.256 &
                {\bf 10.4}   &  1.0   &
                {\bf  2.17}  &  1.01  \\
 G (25.0)     & {\bf  33.3}  &  0.9   &
                {\bf $-$4.43}&  0.34  &
                {\bf 0.0842} & 0.0025 &
                {\bf 29.1}   &  1.4   &
                {\bf  7.394} &  0.214 &
                {\bf 17.7}   &  1.2   &
                {\bf  0.97}  & 0.09   \\
 H (50.0)     & {\bf  22.4}  &  0.7   &
                {\bf $-$5.64}&  0.43  &
                {\bf 0.0789} & 0.0089 &
                {\bf 26.2}   &  3.5   &
                {\bf 12.962} & 0.358  &
                {\bf 25.8}   &  3.1   &
                {\bf  1.15}  &  0.18  \\ \hline
\end{tabular}
\end{center}
\end{table*}

%

%
\begin{table*}
\begin{center}
\begin{tabular}{l|*{6}{r@{ $\pm$}r|}r@{ $\pm$}r}
\multicolumn{10}{l}{\textbf{Series I: $q$ = 0.1, ${\cal N}_{\star}$ = 10,000}} \\ \hline
model ($\mu$) & \multicolumn{2}{c|}{$\langle t_{\rm cc}\rangle $}
              & \multicolumn{2}{c|}{$\langle \Phi_{\rm min} \rangle $}
              & \multicolumn{2}{c|}{$\langle r_{\rm c} \rangle $}
              & \multicolumn{2}{c|}{$\langle {\cal N}_{\rm c} \rangle $}
              & \multicolumn{2}{c|}{$\langle \xi_{\rm min} \rangle $}
              & \multicolumn{2}{c|}{$\langle {\cal N}_{\rm esc} \rangle $}
              & \multicolumn{2}{c}{$\langle E_{\rm esc} \rangle $} \\ \hline
 A (1.25)     & {\bf 2169.9} & 24.7   &      
                {\bf $-$8.35}&  0.52  &      
                {\bf 0.0027} & 0.0003 &      
                {\bf 20.0}   &  2.7   &      
                {\bf  1.000} &  0.000 &      
                {\bf 227.3}  &  3.9   &      
                {\bf 0.40}   & 0.11   \\     
 B (1.5)      & {\bf 1886.8} &  8.6   &
                {\bf $-$7.05}&  0.39  &
                {\bf 0.0042} & 0.0001 &
                {\bf 20.8}   &  1.0   &
                {\bf  1.008} &  0.006 &
                {\bf 172.3}  &  9.3   &
                {\bf 0.35}   & 0.06   \\
 C (2.0)      & {\bf 1218.0} & 20.1   &
                {\bf $-$5.85}&  0.37  &
                {\bf 0.0053} & 0.0007 &
                {\bf 15.8}   &  1.8   &
                {\bf  1.035} &  0.014 &
                {\bf 91.3}   &  4.3   &
                {\bf 0.78}   & 0.17   \\
 D (3.0)      & {\bf 595.1}  &  8.5   &
                {\bf $-$3.97}&  0.17  &
                {\bf 0.0107} & 0.0010 &
                {\bf 15.0}   &  1.9   &
                {\bf  1.170} &  0.024 &
                {\bf 35.3}   & 3.4    &
                {\bf 1.03}   & 0.27   \\
 E (5.0)      & {\bf 277.6}  &  9.7   &
                {\bf $-$4.18}&  0.32  &
                {\bf 0.0235} & 0.0004 &
                {\bf 19.8}   &  1.4   &
                {\bf  1.700} &  0.029 &
                {\bf 19.3}   &  1.4   &
                {\bf  0.63}  &  0.08  \\
 F (10.0)     & {\bf 123.5}  &  2.1   &
                {\bf $-$3.80}&  0.31  &
                {\bf 0.0351} & 0.0055 &
                {\bf 22.3}   &  3.9   &
                {\bf  3.065} &  0.108 &
                {\bf 25.5}   &  2.0   &
                {\bf  1.19}  &  0.04  \\
 G (25.0)     & {\bf  52.4}  &  2.3   &
                {\bf $-$4.92}&  0.42  &
                {\bf 0.0741} & 0.0108 &
                {\bf 36.8}   &  6.1   &
                {\bf  8.040} &  0.440 &
                {\bf 29.0}   &  2.9   &
                {\bf  2.05}  & 0.82   \\
 H (50.0)     & {\bf  37.5}  &  0.6   &
                {\bf $-$6.82}&  0.95  &
                {\bf 0.0894} & 0.0059 &
                {\bf 47.8}   &  8.9   &
                {\bf 14.478} & 0.859  &
                {\bf 42.3}   &  5.3   &
                {\bf  0.93}  &  0.28  \\ \hline
\end{tabular}
\end{center}
\end{table*}


\begin{table*}
\begin{center}
\begin{tabular}{l|*{6}{r@{ $\pm$}r|}r@{ $\pm$}r}
\multicolumn{10}{l}{\textbf{Series I: $q$ = 0.1, ${\cal N}_{\star}$ = 20,000}} \\ \hline
model ($\mu$) & \multicolumn{2}{c|}{$\langle t_{\rm cc}\rangle $}
              & \multicolumn{2}{c|}{$\langle \Phi_{\rm min} \rangle $}
              & \multicolumn{2}{c|}{$\langle r_{\rm c} \rangle $}
              & \multicolumn{2}{c|}{$\langle {\cal N}_{\rm c} \rangle $}
              & \multicolumn{2}{c|}{$\langle \xi_{\rm min} \rangle $}
              & \multicolumn{2}{c|}{$\langle {\cal N}_{\rm esc} \rangle $}
              & \multicolumn{2}{c}{$\langle E_{\rm esc} \rangle $} \\ \hline
 A (1.25)     & {\bf 3859.0} & 00.0   &      
                {\bf $-$8.62}&  0.00  &      
                {\bf 0.0021} & 0.0000 &      
                {\bf 33.0}   &  0.0   &      
                {\bf  1.000} &  0.000 &      
                {\bf 422.0}  &  0.0   &      
                {\bf 0.52}   & 0.00   \\     
 B (1.5)      & {\bf 3318.0} &  0.0   &
                {\bf $-$9.09}&  0.00  &
                {\bf 0.0030} & 0.0000 &
                {\bf 36.0}   &  0.0   &
                {\bf  1.016} &  0.000 &
                {\bf 292.0}  &  0.0   &
                {\bf 0.79}   & 0.00   \\
 C (2.0)      & {\bf 2236.0} &  0.0   &
                {\bf $-$5.64}&  0.00  &
                {\bf 0.0044} & 0.0000 &
                {\bf 23.0}   &  0.0   &
                {\bf  1.081} &  0.000 &
                {\bf 140.0}  &  0.0   &
                {\bf 0.34}   & 0.00   \\
 D (3.0)      & {\bf 1094.0} &  0.0   &
                {\bf $-$3.84}&  0.00  &
                {\bf 0.0114} & 0.0000 &
                {\bf 34.0}   &  0.0   &
                {\bf  1.170} &  0.000 &
                {\bf 55.0}   & 0.0    &
                {\bf 0.56}   & 0.00   \\
 E (5.0)      & {\bf 608.0}  &  0.0   &
                {\bf $-$4.21}&  0.00  &
                {\bf 0.0103} & 0.0000 &
                {\bf 16.0}   &  0.0   &
                {\bf  1.770} &  0.000 &
                {\bf 65.0}   &  0.0   &
                {\bf  2.82}  &  0.00  \\
 F (10.0)     & {\bf 213.0}  &  0.0   &
                {\bf $-$3.70}&  0.00  &
                {\bf 0.0241} & 0.0000 &
                {\bf 24.0}   &  0.0   &
                {\bf  2.950} &  0.000 &
                {\bf 38.0}   &  0.0   &
                {\bf  1.34}  &  0.00  \\
 G (25.0)     & {\bf  78.0}  &  0.0   &
                {\bf $-$4.95}&  0.00  &
                {\bf 0.0863} & 0.0000 &
                {\bf 88.0}   &  0.0   &
                {\bf  7.680} &  0.000 &
                {\bf 30.0}   &  0.0   &
                {\bf  1.41}  & 0.00   \\
 H (50.0)     & {\bf  48.5}  &  0.0   &
                {\bf $-$7.92}&  0.00  &
                {\bf 0.0447} & 0.0000 &
                {\bf  25.0}  &  0.0   &
                {\bf 14.046} & 0.000  &
                {\bf 48.0}   &  0.0   &
                {\bf  0.78}  &  0.00  \\ \hline
\end{tabular}
\end{center}
\end{table*}


%
\begin{table*}
\begin{center}
\begin{tabular}{l|*{6}{r@{ $\pm$}r|}r@{ $\pm$}r}
\multicolumn{9}{l}{\textbf{Series II: $q$ = 0.1, ${\cal N}_{\star}$ = 2500, INS}} \\ \hline
model ($\mu$) & \multicolumn{2}{c|}{$\langle t_{\rm cc}\rangle $}
              & \multicolumn{2}{c|}{$\langle \Phi_{\rm min} \rangle $}
              & \multicolumn{2}{c|}{$\langle r_{\rm c} \rangle $}
              & \multicolumn{2}{c|}{$\langle {\cal N}_{\rm c} \rangle $}
              & \multicolumn{2}{c|}{$\langle \xi_{\rm min} \rangle $}
              & \multicolumn{2}{c|}{$\langle {\cal N}_{\rm esc} \rangle $}
              & \multicolumn{2}{c}{$\langle E_{\rm esc} \rangle $} \\ \hline
 A (1.25)     & {\bf 623.9}  & 11.1   &
                {\bf $-$5.64}&  0.10  &
                {\bf 0.0098} & 0.0005 &
                {\bf 15.2}   & 0.5    &
                \multicolumn{2}{c|}{ --- }&
                {\bf 57.5}   &  2.8   &
                {\bf 0.89}   & 0.22   \\
 B (1.5)      & {\bf 501.5}  &  9.1   &
                {\bf $-$5.05}&  0.10  &
                {\bf 0.0133} & 0.0008 &
                {\bf 14.4}   & 0.8    &
                \multicolumn{2}{c|}{ --- }&
                {\bf 42.0}   &  2.5   &
                {\bf 1.30}   & 0.56   \\
 C (2.0)      & {\bf 304.4}  &  8.0   &
                {\bf $-$4.09}&  0.12  &
                {\bf 0.0225} & 0.0018 &
                {\bf 15.2}   & 1.0    &
                \multicolumn{2}{c|}{ --- }&
                {\bf 16.9}   &  1.1   &
                {\bf 0.35}   & 0.18   \\
 D (3.0)      & {\bf 139.4}  &  4.6   &
                {\bf $-$3.51}&  0.07  &
                {\bf 0.0330} & 0.0023 &
                {\bf 13.5}   & 0.9    &
                \multicolumn{2}{c|}{ --- }&
                {\bf  8.5}   &  0.9   &
                {\bf 0.15}   & 0.11   \\
 E (5.0)      & {\bf  73.6}  &  3.0   &
                {\bf $-$3.51}&  0.08  &
                {\bf 0.0437} & 0.0026 &
                {\bf 12.3}   & 0.8    &
                \multicolumn{2}{c|}{ --- }&
                {\bf  8.4}   &  0.8   &
                {\bf 1.43}   & 0.40   \\
 F (10.0)     & {\bf  32.4}  &  1.7   &
                {\bf $-$3.58}&  0.13  &
                {\bf 0.0722} & 0.0059 &
                {\bf 16.3}   & 1.9    &
                \multicolumn{2}{c|}{ --- }&
                {\bf  5.9}   &  0.8   &
                {\bf 1.01}   & 0.29   \\
 G (25.0)     & {\bf  18.3}  &  0.9   &
                {\bf $-$5.28}&  0.33  &
                {\bf 0.0845} & 0.0070 &
                {\bf 16.3}   & 2.9    &
                \multicolumn{2}{c|}{ --- }&
                {\bf 13.6}   &  1.5   &
                {\bf 0.78}   & 0.11   \\
 H (50.0)     & {\bf  15.2}  &  0.9   &
                {\bf $-$5.52}&  0.27  &
                {\bf 0.0892} & 0.0050 &
                {\bf 13.9}   & 1.4    &
                \multicolumn{2}{c|}{ --- }&
                {\bf 25.7}   &  2.2   &
                {\bf 0.82}   & 0.10   \\ \hline
\end{tabular}
\end{center}
\end{table*}

%

%
\begin{table*}
\begin{center}
\begin{tabular}{l|*{6}{r@{ $\pm$}r|}r@{ $\pm$}r}
\multicolumn{9}{l}{\textbf{Series II: $q$ = 0.1, ${\cal N}_{\star}$ = 2500, OUT}} \\ \hline
model ($\mu$) & \multicolumn{2}{c|}{$\langle t_{\rm cc}\rangle $}
              & \multicolumn{2}{c|}{$\langle \Phi_{\rm min} \rangle $}
              & \multicolumn{2}{c|}{$\langle r_{\rm c} \rangle $}
              & \multicolumn{2}{c|}{$\langle {\cal N}_{\rm c} \rangle $}
              & \multicolumn{2}{c|}{$\langle \xi_{\rm min} \rangle $}
              & \multicolumn{2}{c|}{$\langle {\cal N}_{\rm esc} \rangle $}
              & \multicolumn{2}{c}{$\langle E_{\rm esc} \rangle $} \\ \hline
 A (1.25)     & {\bf 673.7}  &  8.6   &
                {\bf $-$6.08}& 0.16   &
                {\bf 0.0088} & 0.0004 &
                {\bf 14.8}   & 0.6    &
                \multicolumn{2}{c|}{ --- }&
                {\bf 61.4}   & 2.5    &
                {\bf 1.03}   & 0.41   \\
 B (1.5)      & {\bf 646.1}  & 11.3   &
                {\bf $-$5.12}&  0.14  &
                {\bf 0.0129} & 0.0008 &
                {\bf 17.1}   & 0.9    &
                \multicolumn{2}{c|}{ --- }&
                {\bf 50.1}   &  2.5   &
                {\bf 0.66}   & 0.14   \\
 C (2.0)      & {\bf 508.5}  & 10.8   &
                {\bf $-$4.16}&  0.12  &
                {\bf 0.0227} & 0.0014 &
                {\bf 18.2}   & 1.2    &
                \multicolumn{2}{c|}{ --- }&
                {\bf 32.6}   &  1.6   &
                {\bf 1.77}   & 0.42   \\
 D (3.0)      & {\bf 304.1}  &  7.3   &
                {\bf $-$3.23}&  0.05  &
                {\bf 0.0476} & 0.0026 &
                {\bf 24.8}   & 1.8    &
                \multicolumn{2}{c|}{ --- }&
                {\bf 13.0}   &  0.8   &
                {\bf 1.10}   & 0.30   \\
 E (5.0)      & {\bf 153.7}  &  4.7   &
                {\bf $-$2.98}&  0.06  &
                {\bf 0.0756} & 0.0042 &
                {\bf 29.6}   & 2.8    &
                \multicolumn{2}{c|}{ --- }&
                {\bf  8.5}   &  0.6   &
                {\bf 1.40}   & 0.32   \\
 F (10.0)     & {\bf  92.7}  &  3.0   &
                {\bf $-$3.23}&  0.08  &
                {\bf 0.0892} & 0.0033 &
                {\bf 23.8}   & 1.6    &
                \multicolumn{2}{c|}{ --- }&
                {\bf 10.0}   &  0.9   &
                {\bf 2.17}   & 0.47   \\
 G (25.0)     & {\bf  52.5}  &  1.5   &
                {\bf $-$4.58}&  0.27  &
                {\bf 0.0857} & 0.0037 &
                {\bf 17.5}   & 1.2    &
                \multicolumn{2}{c|}{ --- }&
                {\bf 17.0}   &  1.6   &
                {\bf 1.76}   & 0.26   \\
 H (50.0)     & {\bf  47.6}  &  2.3   &
                {\bf $-$6.86}&  0.29  &
                {\bf 0.0664} & 0.0032 &
                {\bf  9.7}   & 0.7    &
                \multicolumn{2}{c|}{ --- }&
                {\bf 32.0}   &  2.8   &
                {\bf 1.94}   & 0.32   \\ \hline
\end{tabular}
\end{center}
\end{table*}
%


%
\begin{table*}
\begin{center}
\begin{tabular}{l|*{6}{r@{ $\pm$}r|}r@{ $\pm$}r}
\multicolumn{10}{l}{\textbf{Series III: $q$ = 0.05, ${\cal N}_{\star}$ = 2500}} \\ \hline
model ($\mu$) & \multicolumn{2}{c|}{$\langle t_{\rm cc}\rangle $}
              & \multicolumn{2}{c|}{$\langle \Phi_{\rm min} \rangle $}
              & \multicolumn{2}{c|}{$\langle r_{\rm c} \rangle $}
              & \multicolumn{2}{c|}{$\langle {\cal N}_{\rm c} \rangle $}
              & \multicolumn{2}{c|}{$\langle \xi_{\rm min} \rangle $}
              & \multicolumn{2}{c|}{$\langle {\cal N}_{\rm esc} \rangle $}
              & \multicolumn{2}{c}{$\langle E_{\rm esc} \rangle $} \\ \hline
 K (1.25)     & {\bf 689.3}  &  7.8   &      
                {\bf $-$5.88}&  0.21  &      
                {\bf 0.0094} & 0.0005 &      
                {\bf 16.0}   &  0.6   &      
                {\bf  1.000} &  0.000 &      
                {\bf 60.7}   &  2.5   &      
                {\bf 0.53}   & 0.12   \\     
 L (1.5)      & {\bf 605.3}  & 11.8   &
                {\bf $-$5.19}&  0.19  &
                {\bf 0.0140} & 0.0009 &
                {\bf 17.2}   &  0.8   &
                {\bf  1.001} &  0.000 &
                {\bf 51.0}   &  2.7   &
                {\bf 0.61}   & 0.16   \\
 M (2.0)      & {\bf 428.8}  &  9.8   &
                {\bf $-$3.89}&  0.11  &
                {\bf 0.0264} & 0.0022 &
                {\bf 19.2}   &  1.4   &
                {\bf  1.009} &  0.008 &
                {\bf 25.8}   &  1.7   &
                {\bf 0.82}   & 0.28   \\
 N (3.0)      & {\bf 219.3}  &  7.4   &
                {\bf $-$3.20}&  0.06  &
                {\bf 0.0443} & 0.0028 &
                {\bf 19.5}   &  1.6   &
                {\bf  1.092} &  0.034 &
                {\bf  9.3}   & 1.1    &
                {\bf 0.85}   & 0.27   \\
 P (5.0)      & {\bf  96.1}  &  4.5   &
                {\bf $-$3.07}&  0.06  &
                {\bf 0.0672} & 0.0046 &
                {\bf 23.6}   &  2.4   &
                {\bf  1.510} &  0.034 &
                {\bf  4.3}   &  0.7   &
                {\bf 1.60}   & 0.54   \\
 Q (10.0)     & {\bf  44.9}  &  1.4   &
                {\bf $-$3.34}&  0.11  &
                {\bf 0.0896} & 0.0053 &
                {\bf 23.5}   &  2.4   &
                {\bf  2.638} &  0.068 &
                {\bf  4.2}   &  0.6   &
                {\bf 1.06}   & 0.19   \\
 R (25.0)     & {\bf  27.6}  &  1.2   &
                {\bf $-$4.53}&  0.21  &
                {\bf 0.0903} & 0.0046 &
                {\bf 19.0}   &  1.9   &
                {\bf  4.348} &  0.290 &
                {\bf  9.0}   &  1.4   &
                {\bf 1.37}   & 0.44   \\ \hline
\end{tabular}
\end{center}
\end{table*}


%
\begin{table*}
\begin{center}
\begin{tabular}{l|*{6}{r@{ $\pm$}r|}r@{ $\pm$}r}
\multicolumn{10}{l}{\textbf{Series IV: $q$ = 0.2, ${\cal N}_{\star}$ = 2500}} \\ \hline
model ($\mu$) & \multicolumn{2}{c|}{$\langle t_{\rm cc}\rangle $}
              & \multicolumn{2}{c|}{$\langle \Phi_{\rm min} \rangle $}
              & \multicolumn{2}{c|}{$\langle r_{\rm c} \rangle $}
              & \multicolumn{2}{c|}{$\langle {\cal N}_{\rm c} \rangle $}
              & \multicolumn{2}{c|}{$\langle \xi_{\rm min} \rangle $}
              & \multicolumn{2}{c|}{$\langle {\cal N}_{\rm esc} \rangle $}
              & \multicolumn{2}{c}{$\langle E_{\rm esc} \rangle $} \\ \hline
 T (1.25)     & {\bf 657.1}  & 11.0   &      
                {\bf $-$5.88}&  0.17  &      
                {\bf 0.0102} & 0.0006 &      
                {\bf 15.2}   &  0.9   &      
                {\bf  1.001} &  0.000 &      
                {\bf 62.5}   &  2.6   &      
                {\bf 0.81}   & 0.21   \\     
 U (1.5)      & {\bf 549.4}  &  7.6   &
                {\bf $-$5.00}&  0.14  &
                {\bf 0.0134} & 0.0009 &
                {\bf 14.6}   &  0.8   &
                {\bf  1.000} &  0.002 &
                {\bf 49.5}   &  2.7   &
                {\bf 1.17}   & 0.28   \\
 V (2.0)      & {\bf 343.8}  &  9.5   &
                {\bf $-$4.51}&  0.11  &
                {\bf 0.0166} & 0.0011 &
                {\bf 12.3}   &  0.6   &
                {\bf  1.067} &  0.012 &
                {\bf 26.7}   &  2.2   &
                {\bf 1.04}   & 0.35   \\
 W (3.0)      & {\bf 174.0}  &  3.9   &
                {\bf $-$3.87}&  0.11  &
                {\bf 0.0244} & 0.0020 &
                {\bf 11.8}   &  0.8   &
                {\bf  1.334} &  0.016 &
                {\bf 16.9}   & 1.3    &
                {\bf 0.61}   & 0.04   \\
 X (5.0)      & {\bf  86.3}  &  2.2   &
                {\bf $-$3.55}&  0.09  &
                {\bf 0.0391} & 0.0028 &
                {\bf 12.0}   &  0.9   &
                {\bf  2.069} &  0.059 &
                {\bf 13.9}   &  0.8   &
                {\bf 0.81}   & 0.08   \\
 Y (10.0)     & {\bf  48.6}  &  2.4   &
                {\bf $-$3.97}&  0.16  &
                {\bf 0.0564} & 0.0041 &
                {\bf 11.9}   &  1.0   &
                {\bf  4.063} &  0.161 &
                {\bf 21.1}   &  1.6   &
                {\bf 1.06}   & 0.11   \\
 Z (25.0)     & {\bf  22.7}  &  1.0   &
                {\bf $-$4.72}&  0.24  &
                {\bf 0.0869} & 0.0049 &
                {\bf 15.5}   &  1.6   &
                {\bf  9.676} &  0.267 &
                {\bf 30.1}   &  1.97  &
                {\bf 0.74}   & 0.07   \\ \hline
\end{tabular}
\end{center}
\end{table*}


%
\begin{table*}
\begin{center}
\begin{tabular}{l|*{6}{r@{ $\pm$}r|}r@{ $\pm$}r}
\multicolumn{10}{l}{\textbf{Series V: $q$ = 0.4, ${\cal N}_{\star}$ = 2500}} \\ \hline
model ($\mu$) & \multicolumn{2}{c|}{$\langle t_{\rm cc}\rangle $}
              & \multicolumn{2}{c|}{$\langle \Phi_{\rm min} \rangle $}
              & \multicolumn{2}{c|}{$\langle r_{\rm c} \rangle $}
              & \multicolumn{2}{c|}{$\langle {\cal N}_{\rm c} \rangle $}
              & \multicolumn{2}{c|}{$\langle \xi_{\rm min} \rangle $}
              & \multicolumn{2}{c|}{$\langle {\cal N}_{\rm esc} \rangle $}
              & \multicolumn{2}{c}{$\langle E_{\rm esc} \rangle $} \\ \hline
T$\prime$ (1.25)&{\bf 639.7}  &  9.9   &      
                {\bf $-$5.79}&  0.16  &      
                {\bf 0.0092} & 0.0007 &      
                {\bf 14.0}   &  0.8   &      
                {\bf  1.000} &  0.000 &      
                {\bf 63.0}   &  2.4   &      
                {\bf 0.68}   & 0.13   \\     
U$\prime$ (1.5)&{\bf 523.2}  &  8.2   &
                {\bf $-$5.24}&  0.12  &
                {\bf 0.0117} & 0.0006 &
                {\bf 14.4}   &  0.8   &
                {\bf  1.007} &  0.003 &
                {\bf 51.4}   &  2.7   &
                {\bf 0.81}   & 0.22   \\
V$\prime$ (2.0)&{\bf 372.3}  & 10.3   &
                {\bf $-$4.95}&  0.15  &
                {\bf 0.0149} & 0.0007 &
                {\bf 13.3}   &  0.6   &
                {\bf  1.094} &  0.013 &
                {\bf 42.3}   &  2.9   &
                {\bf 0.85}   & 0.24   \\
W$\prime$ (3.0)&{\bf 212.8}  &  4.3   &
                {\bf $-$4.27}&  0.10  &
                {\bf 0.0195} & 0.0014 &
                {\bf 11.3}   &  0.6   &
                {\bf  1.436} &  0.039 &
                {\bf 38.7}   & 2.4    &
                {\bf 0.57}   & 0.03   \\
X$\prime$ (5.0)&{\bf 130.7}  &  4.0   &
                {\bf $-$4.19}&  0.14  &
                {\bf 0.0274} & 0.0015 &
                {\bf 10.4}   &  0.6   &
                {\bf  2.200} &  0.047 &
                {\bf 43.7}   &  2.9   &
                {\bf 0.81}   & 0.05   \\
Y$\prime$ (10.0)&{\bf  66.8} &  2.5   &
                {\bf $-$4.26}&  0.12  &
                {\bf 0.0440} & 0.0026 &
                {\bf  9.9}   &  0.5   &
                {\bf  4.316} &  0.087 &
                {\bf 50.9}   &  3.5   &
                {\bf 0.88}   & 0.05   \\ \hline
\end{tabular}
\end{center}
\end{table*}

\section{$N$-body--units and timescales}   \label{ch:nbtime}

Dimensionless units, so--called ``$N$-body units'',
were used throughout the calculations.
They are obtained when setting the gravitational
constant $G$ and the initial total cluster mass ${\cal M}_{\rm cl}$
equal to 1, and the initial total energy $E$ to $-1/4$
\citep{HM86,AHW74}.
Since the total energy $E$ of the system is $E = K + W$
with $K = \frac{1}{2}{\cal M}_{\rm cl}\langle v^2\rangle$ being
the total kinetic energy and $W = -(3\pi/32)G{\cal M}_{\rm cl}^2/R$
the potential energy of the Plummer sphere, we
find from the virial theorem that
\begin{eqnarray}
   E = \frac{1}{2}W = -\frac{3\pi}{64}\frac{G{\cal M}_{\rm cl}^2}{R}.
           \label{eq:totenergy}
\end{eqnarray}
$R$ is a quantity which determines the length scale
of a Plummer sphere.
Using the specific definitions for $G$, ${\cal M}_{\rm cl}$, and $E$ above,
this scaling radius becomes $R = 3\pi/16$ in dimensionless
units.
The half mass radius $r_{\rm h}$ can easily be evaluated
by the formula (e.g. \citealt{Spitzer87}):
\begin{eqnarray}
   M(r) = {\cal M}_{\rm cl}\frac{r^3/R^3}{(1+r^2/R^2)^{3/2}}
\end{eqnarray}
when setting $M(r_{\rm h}) = \frac{1}{2}{\cal M}_{\rm cl}$.
It yields $r_{\rm h} = (2^{2/3}-1)^{-1/2}R = 1.30\, R$.
The half--mass radius is located at the scale
length of $R = 0.766$,
or about 3/4 ``$N$-body--radii''.

The initial half--mass crossing time of a particle is
\begin{eqnarray}
  t_{\rm cr} = \frac{ G{\cal M}_{\rm cl}^{5/2} }{ (2E)^{3/2} }.
\end{eqnarray}
Since the $N$-body--unit of time, $t_{\rm NB}$, is 1 when
\begin{eqnarray}
  t_{\rm NB} = \frac{ G{\cal M}_{\rm cl}^{5/2} }{ (-4E)^{3/2} }, \label{eq:tnb}
\end{eqnarray}
immediately follows that $t_{\rm cr}/t_{\rm NB} = 2\sqrt{2}$.

In the situations considered here, the evolution of the cluster is
driven by 2-body relaxation. Therefore, a natural time scale is the
{\em half-mass relaxation time}. We use the definition of
\citet{Spitzer87}, 
\be
\label{eq_rel_time}
T_{\rm rh} = \frac{0.138 N}{\ln \Lambda}
\left(\frac{R_{1/2}^3}{G{\cal M}_{\rm cl}}\right)^{1/2}.
\ee
For a Plummer model, the half-mass radius is $R_{1/2}
= 1.305\, R$. 
${\cal M}_{\rm cl}$ is the total stellar mass.

\bibliographystyle{mn}
\bibliography{aamnem99,biblio_tesi}

\begin{thebibliography}{45}
\expandafter\ifx\csname natexlab\endcsname\relax\def\natexlab#1{#1}\fi

\bibitem[{{Aarseth} {et~al.}(1974){Aarseth}, {Henon}, \& {Wielen}}]{AHW74}
{Aarseth} S.~J., {Henon} M., {Wielen} R., 1974, A\&A, 37, 183

\bibitem[{{Baumgardt}(2001)}]{Baumgardt01}
{Baumgardt} H., 2001, MNRAS, 325, 1323

\bibitem[{{Baumgardt} {et~al.}(2002){Baumgardt}, {Hut}, \&
  {Heggie}}]{BaumgardtEtAl02}
{Baumgardt} H., {Hut} P., {Heggie} D.~C., 2002, MNRAS, 336, 1069

\bibitem[{{Bettwieser} \& {Inagaki}(1985)}]{BI85}
{Bettwieser} E., {Inagaki} S., 1985, MNRAS, 213, 473

\bibitem[{{Bonnell} \& {Davies}(1998)}]{BD98}
{Bonnell} I.~A., {Davies} M.~B., 1998, MNRAS, 295, 691

\bibitem[{{Cohn}(1980)}]{Cohn80}
{Cohn} H., 1980, ApJ, 242, 765

\bibitem[{{Drukier} {et~al.}(1999){Drukier}, {Cohn}, {Lugger}, \&
  {Yong}}]{DCLY99}
{Drukier} G.~A., {Cohn} H.~N., {Lugger} P.~M., {Yong} H., 1999, ApJ, 518, 233

\bibitem[{{Fregeau} {et~al.}(2002){Fregeau}, {Joshi}, {Portegies Zwart}, \&
  {Rasio}}]{FregeauEtAl02}
{Fregeau} J.~M., {Joshi} K.~J., {Portegies Zwart} S.~F., {Rasio} F.~A., 2002,
  apj, 570, 171

\bibitem[{{Giersz} \& {Heggie}(1994{\natexlab{a}})}]{GH94a}
{Giersz} M., {Heggie} D.~C., 1994{\natexlab{a}}, MNRAS, 268, 257

\bibitem[{{Giersz} \& {Heggie}(1994{\natexlab{b}})}]{GH94b}
---, 1994{\natexlab{b}}, MNRAS, 270, 298

\bibitem[{{Giersz} \& {Heggie}(1996)}]{GH96}
---, 1996, MNRAS, 279, 1037

\bibitem[{{Goodman}(1984)}]{Goodman84}
{Goodman} J., 1984, ApJ, 280, 298

\bibitem[{{Goodman}(1987)}]{Goodman87}
---, 1987, apj, 313, 576

\bibitem[{{H{\' e}non}(1965)}]{Henon65}
{H{\' e}non} M., 1965, Annales d'Astrophysique, 28, 62

\bibitem[{{Heggie} \& {Mathieu}(1986)}]{HM86}
{Heggie} D.~C., {Mathieu} R.~D., 1986, in The Use of Supercomputers in Stellar
  Dynamics, {Hut} P., {McMillan} S. L.~W., eds., Springer-Verlag, p. 233

\bibitem[{{H{\'e}non}(1969)}]{Henon69}
{H{\'e}non} M., 1969, A\&A, 2, 151

\bibitem[{{H{\'{e}}non}(1973)}]{Henon73}
{H{\'{e}}non} M., 1973, in Dynamical structure and evolution of stellar
  systems, Lectures of the 3rd Advanced Course of the Swiss Society for
  Astronomy and Astrophysics (SSAA), {Martinet} L., {Mayor} M., eds., pp.
  183--260

\bibitem[{{H{\'e}non}(1975)}]{Henon75}
{H{\'e}non} M., 1975, in IAU Symp. 69: Dynamics of Stellar Systems, {Hayli} A.,
  ed., p. 133

\bibitem[{{Hillenbrand} \& {Hartmann}(1998)}]{HH98}
{Hillenbrand} L.~A., {Hartmann} L.~W., 1998, apj, 492, 540

\bibitem[{{Inagaki} \& {Wiyanto}(1984)}]{IW84}
{Inagaki} S., {Wiyanto} P., 1984, PASJ, 36, 391

\bibitem[{{Joshi} {et~al.}(2000){Joshi}, {Rasio}, \& {Portegies
  Zwart}}]{JRPZ00}
{Joshi} K.~J., {Rasio} F.~A., {Portegies Zwart} S., 2000, ApJ, 540, 969

\bibitem[{{Kim} {et~al.}(1998){Kim}, {Lee}, \& {Goodman}}]{KLG98}
{Kim} S.~S., {Lee} H.~M., {Goodman} J., 1998, apj, 495, 786

\bibitem[{{Lee}(1995)}]{Lee95}
{Lee} H.~M., 1995, MNRAS, 272, 605

\bibitem[{{Lightman} \& {Fall}(1978)}]{LF78}
{Lightman} A.~P., {Fall} S.~M., 1978, apj, 221, 567

\bibitem[{{Makino}(1996)}]{Makino96}
{Makino} J., 1996, ApJ, 471, 796

\bibitem[{{Marchant} \& {Shapiro}(1980)}]{MS80}
{Marchant} A.~B., {Shapiro} S.~L., 1980, ApJ, 239, 685

\bibitem[{{McCaughrean} \& {Stauffer}(1994)}]{McMS94}
{McCaughrean} M.~J., {Stauffer} J.~R., 1994, AJ, 108, 1382

\bibitem[{{McMillan} {et~al.}(1981){McMillan}, {Lightman}, \& {Cohn}}]{McMLC81}
{McMillan} S. L.~W., {Lightman} A.~P., {Cohn} H., 1981, ApJ, 251, 436

\bibitem[{{Meylan} \& {Heggie}(1997)}]{MH97}
{Meylan} G., {Heggie} D.~C., 1997, A\&AR, 8, 1

\bibitem[{{Portegies Zwart} \& {McMillan}(2000)}]{PZMM00}
{Portegies Zwart} S.~F., {McMillan} S.~L.~W., 2000, apjl, 528, L17

\bibitem[{{Quinlan}(1996)}]{Quinlan96}
{Quinlan} G.~D., 1996, New Astronomy, 1, 255

\bibitem[{{Raboud} \& {Mermilliod}(1998)}]{RM98}
{Raboud} D., {Mermilliod} J.-C., 1998, A\&A, 333, 897

\bibitem[{{Shapiro}(1977)}]{Shapiro77}
{Shapiro} S.~L., 1977, ApJ, 217, 281

\bibitem[{{Spitzer}(1987)}]{Spitzer87}
{Spitzer} L., 1987, Dynamical evolution of globular clusters. Princeton
  University Press

\bibitem[{{Spitzer} \& {Shull}(1975)}]{SS75a}
{Spitzer} L., {Shull} J.~M., 1975, ApJ, 200, 339

\bibitem[{{Spitzer}(1969)}]{Spitzer69}
{Spitzer} L.~J., 1969, ApJ Lett., 158, 139

\bibitem[{{Spitzer} \& {Hart}(1971)}]{SH71b}
{Spitzer} L.~J., {Hart} M.~H., 1971, ApJ, 166, 483

\bibitem[{{Spurzem} \& {Aarseth}(1996)}]{SA96}
{Spurzem} R., {Aarseth} S.~J., 1996, MNRAS, 282, 19

\bibitem[{{Spurzem} \& {Takahashi}(1995)}]{ST95}
{Spurzem} R., {Takahashi} K., 1995, MNRAS, 272, 772

\bibitem[{{Stodo{\l}kiewicz}(1982)}]{Stodol82}
{Stodo{\l}kiewicz} J.~S., 1982, Acta Astron., 32, 63

\bibitem[{{Takahashi}(1993)}]{Takahashi93}
{Takahashi} K., 1993, PASJ, 45, 233

\bibitem[{{Takahashi}(1995)}]{Takahashi95}
---, 1995, PASJ, 47, 561

\bibitem[{{von Hoerner}(1960)}]{Hoerner60}
{von Hoerner} S., 1960, Zeitschrift fur Astrophysics, 50, 184

\bibitem[{{Watters} {et~al.}(2000){Watters}, {Joshi}, \& {Rasio}}]{WJR00}
{Watters} W.~A., {Joshi} K.~J., {Rasio} F.~A., 2000, ApJ, 539, 331

\bibitem[{{Wielen}(1975)}]{Wielen75}
{Wielen} R., 1975, in IAU Symp. 69: Dynamics of the Solar Systems, pp. 119--131

\end{thebibliography}
\label{lastpage}
\end{document}